\documentclass[preprint,12pt]{elsarticle}
\usepackage{graphicx}
\usepackage{amssymb}
\usepackage{amsthm}
\usepackage{multirow}
\usepackage{array}
\usepackage[caption=false,font=footnotesize]{subfig}
\usepackage{url}
\usepackage{cases}
\graphicspath{{figs/}}
\DeclareGraphicsExtensions{.eps}

\newcommand{\eat}[1]{}
\newtheorem{remark}{Remark}

\begin{document}
\begin{frontmatter}


%
%
%
%
%
\title{Heterogeneous download times in bandwidth-homogeneous BitTorrent swarms
}

\author[label1]{Fabricio Murai}
\author[label2]{Antonio A. de A. Rocha}
\author[label1]{\\ Daniel R. Figueiredo}
\author[label1]{Edmundo A. de Souza e Silva}


\address[label1]{COPPE/Systems Engineering and Computer Science Program,\\
Federal University of Rio de Janeiro,\\
Rio de Janeiro, Brazil\\
\{fabricio,daniel,edmundo\}@land.ufrj.br\\
\ }
\address[label2]{Computer Science Department,\\
Fluminense Federal University,\\
Niter\'oi, Brazil\\
arocha@ic.uff.br}




\begin{abstract}
Modeling and understanding BitTorrent (BT) dynamics is a recurrent research
topic mainly due to its high complexity and tremendous practical efficiency.
Over the years, different models have uncovered various phenomena exhibited by
the system, many of which have direct impact on its performance. In this paper
we identify and characterize a phenomenon that has not been previously observed:
homogeneous peers (with respect to their upload capacities) experience
heterogeneous download times. This behavior has direct impact on peer and system
performance, such as high variability of download times, unfairness with respect
to peer arrival order, bursty departures and content synchronization. Detailed
packet-level simulations and prototype-based experiments on the Internet were
performed to characterize this phenomenon. We also develop a mathematical model
that accurately predicts the heterogeneous download rates of the homogeneous
peers as a function of their content. In addition, we apply the model to
calculate lower and upper bounds to  the number of departures that occur in a
burst. The heterogeneous download rates are more prevalent in unpopular swarms
(very few peers). Although few works have addressed this kind of swarm, these 
by far represent the most common type of swarm in BT.\footnote{Published in Computer Networks \cite{murai2012comnet}.} 
\end{abstract}

\begin{keyword}
Peer-to-peer \sep BitTorrent \sep Performance evaluation \sep Modeling
\end{keyword}

\end{frontmatter}

\section{Introduction}

Peer-to-peer (P2P) applications have widely been used for content recovery in
Internet. Among them, BitTorrent (BT)~\cite{bt} is one of the most popular, used
by millions daily to retrieve millions of files (movies, TV series, music, etc),
accounting for large fractions of today's Internet traffic~\cite{urlinternet}.
The mainstream success of BT is closely related to its performance (e.g., fast
download times) and together with its high complexity, has triggered the
interest of researchers. 

Understanding and characterizing the performance of BT through mathematical models has 
been an active topic of research~\cite{xia_muppala_2010}.
Several studies have uncovered peculiar aspects BT's dynamic, many of which have 
direct impact on system performance. 
Moreover, models that capture user and system performance under 
homogeneous and heterogeneous peer population (with respect to their upload capacities) 
have been proposed for various scenarios \cite{yang_veciana_2004,qiu_srikant_2004,liao_papadopoulos_psounis_2007,chow_golubchik_misra_2009}.
However, most proposed models target large-scale systems, either with a large and 
fixed initial peer population or relatively high peer arrival rates.

We consider a BT swarm where all peers have identical upload capacities but unconstrained 
(or large) download capacities. In this context, we identify and characterize a phenomenon 
that has not been previously observed: homogeneous peers experience heterogeneous 
download rates. Although this is expected in swarms where peers have
different capacities, in homogeneous swarms, peers should, at first,
exhibit similar 
average performance. Thus, we focus in the latter type of swarm, for which the
described behavior has not been captured by any prior model 
(to the best of our knowledge). Moreover, this observation has several important 
implications, such as high variability of download times, unfairness with respect to 
peer arrival order, bursty departures and content synchronization among the peers. 
Two peers are said to be content-synchronized after their content become identical at
a given instant. This last consequence is particularly critical since it is closely related to 
the missing piece syndrome \cite{mathieu2,hajek_zhu_2010,hajek2}, a scenario where a very large
number of peers have all except a single missing piece.

We characterize the fact that homogeneous peers experience heterogeneous download rates and 
its various consequences by using detailed packet-level simulations and prototype-based 
experiments on the Internet. To underpin critical parameters for this behavior, we consider 
various scenarios. We show that peer arrival times strongly influence their
average download rate. We also develop a mathematical model that explains the phenomenon and predicts 
the heterogeneous download rates of the homogeneous peers as a function of their content. 
The comparison of model predictions with simulation results indicate the model is quite 
accurate. More importantly, the model sheds light on the key insight for this behavior: 
upload capacity allocation of peers in BT depends fundamentally on piece interest relationship, 
which for unpopular swarms can be rather asymmetric. We also apply the model to calculate lower and upper bounds to
the number of departures that occur in a burst.

\noindent{\bf Remark: The case for unpopular swarms with seeds}

The phenomenon we identify is more prevalent in swarms that have a very small 
peer population and a single seed (peer with entire content) with limited bandwidth. 
However, this is by far the most prevalent kind of swarm in BT, as observed by different and 
independent measurement studies. In particular, it has been shown that inter-arrival times of peers into 
swarms increase exponentially with the age of the swarm \cite{guo,Kaune2010}. Thus, some time after 
it has been created, swarms receive few peers and therefore have a very small size. A detailed 
measurement study of swarm sizes in BT considering various repositories 
and various media types has also recently appeared in the literature \cite{Hossfeld2011}. Their 
results indicate that 70\% of active swarms from different repositories have less than 10 
peers (Figure 2 in \cite{Hossfeld2011}). When considering swarms that do not change size over a 
relative short time, 97\% of them have less than 5 peers (Figure 3 in \cite{Hossfeld2011}). 

\begin{figure}[!t]
\begin{center}
\captionsetup[subfloat]{}
\subfloat[Empirical complementary cumulative distribution of swarm sizes with at least one seed for different days (each curve corresponds to a day).]{\label{fig:size-distrib-1}
\includegraphics[width=2.6in]{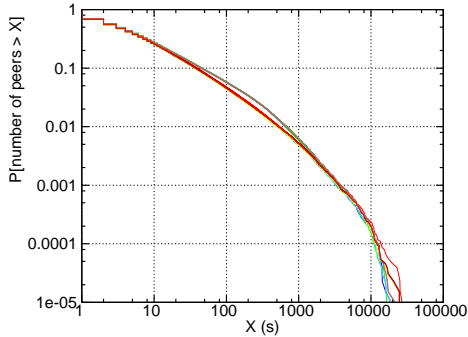}}
\hspace{1cm}
\subfloat[Fraction of swarms of size $K$ with at least one seed in Torlock.com for different days and swarm sizes (each bar corresponds to a day).]{\label{fig:crawling_frac_k}
\includegraphics[width=3.5in]{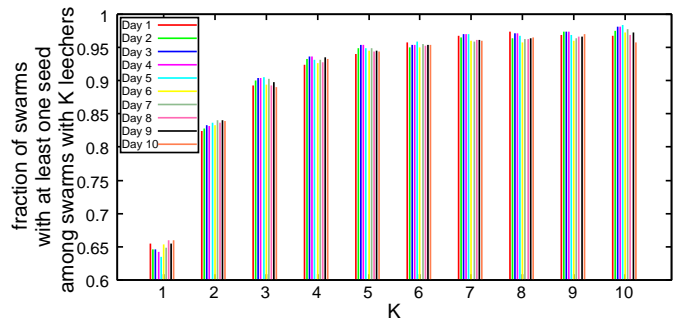}}
\caption{\label{fig:size-distrib} Distribution of swarm sizes and fraction of swarms with at least one seed in Torlock.com for ten consecutive measurement days.}
\end{center}
\end{figure}

We have also conducted measurements in Torlock.com, one of the most popular Torrent Search Engines 
available in the Internet nowadays. In particular, we collected information concerning {\it swarm health} 
(number of peers, number of seeds, etc) on all available swarms in the website (around 150,000) 
once a day for ten consecutive days in November 2011. Each swarm has a size which is given by the number 
of peers connected to the swarm (seeders plus leechers) at the time data was collected. 
Figure \ref{fig:size-distrib-1} shows the empirical complementary cumulative distribution of swarm 
sizes for all ten days, considering only swarms that have at least one seed (around 130,000 swarms). 
Interestingly, swarm size distribution is heavy tailed, with some swarms having a size 
1000 times larger than the average. Moreover, most swarms are very small: about 58\% of the swarms have 
less than 5 peers and about 73\% have less than 10 peers. Finally, this observation is persistent and 
consistent over the ten measurement days, indicating that small swarms are very prevalent in BT. 
Intuitively, swarms without any seeds are not likely to exist in BT since the content 
may not be fully available in them. Figure \ref{fig:crawling_frac_k} shows the fraction of swarms 
of size $K$ with at least one seed. As expected, the fraction of swarms with at least one seed is very 
large, more than 90\% for all swarm sizes greater than 2. Moreover, as the size of the swarm increases, 
this fraction also increases. Again, we observe that this is consistent over the ten measurement days, 
indicating that swarms with at least one seed are very frequent, even when considering unpopular swarms, 
with sizes less than 5. 

Finally, as supported by experimental evidence, unpopular swarms (swarms of very small sizes, e.g., five or less 
peers) with at least one seed are very common in the real world. Thus, they are the focus point of this paper, 
although we will present and discuss some generalizations.

The rest of this paper is organized as follows. In~\textsection\ref{sec:bt} we present 
a brief overview of BT and motivate the phenomenon we have identified.
In~\textsection\ref{sec:problem} we characterize the phenomenon and its consequences using 
simulations and experiments with a real BT application. \textsection\ref{sec:model} presents 
our mathematical model, its validation in comparison with simulations, and some model 
generalizations. In \textsection\ref{sec:application} we apply 
the model to make predictions about bursty departures. We include a discussion
and possible model 
extensions as well as present some related work in \textsection\ref{sec:disc} and \textsection\ref{sec:related}, 
respectively. Finally, we conclude the paper in~\textsection\ref{sec:conc}.

\section{BT overview and the observed behavior}
\label{sec:bt}

\subsection{Brief BT overview}

BT is a swarm based file sharing P2P application. Swarm is a set of users
(peers) interested in downloading and/or sharing the same content (a single or a
bundle of files). The content is chopped into pieces (chunks) which are
exchanged among peers connected to the swarm. The entities in a swarm may be of
three different types: (i) the seeds which are peers that have a complete copy
of the content and are still connected to the system altruistically uploading
data to other peers; (ii) the leechers which are peers that have not yet fully
recovered the content and are actively downloading and simultaneously uploading
the chunks; and, (iii) the tracker which is a kind of swarm coordinator, it
keeps track of the leechers and seeds connected to the swarm.

Periodically, the tracker distributes lists with a random subset of peers
connected to the swarm to promote the interaction among participating peers. In
a first interaction, two peers exchange their bitmaps (a list of all file chunks
they have downloaded). All updates in their bitmaps are
reported by the leecher to its neighbors. 

In order to receive new chunks, the leecher must send ``Interested'' messages to
all peers that announced to have the wanted pieces in their respective bitmaps.
Because of the rarest first approach specified in BT protocol, leechers
prioritize to download first the chunks that are scarcer in the swarm. Once a
sub-piece of any chunk is received, the ``strict priority'' policy defines that
the remaining sub-pieces from that particular chunk must be requested before
starting the download of any other chunk.

Whenever an ``Interested'' message is received, peers have to decide whether to
``unchoke'' that leecher and serve the piece or to ``choke'' the peer and ignore
the request. Leechers preferentially upload content to other leechers that
reciprocate likewise, it is based on a ``tit-for-tat'' incentive strategy defined
by BT's protocol. More precisely, a major fraction of its bandwidth is allocated
to serve the peers that have contributed the most to the leecher.
However, a minor fraction of its bandwidth must be dedicated to
altruistically serve leechers that have never reciprocated. This policy,
referred to
as ``optimistic unchoke'', is useful for leechers to bootstrap new reciprocity
relationships. As the seeds do not reciprocate, they adopt the ``optimistic
unchoke'' approach all the time. These BT policies were designed with the main
purpose of giving all leechers a ``fair share'' of bandwidth. It means that
peers uploading in higher rates will receive in higher download rates, and in a
population of leechers uploading at the same rate, they all must reach equal download rates.

\subsection{The observed behavior}
\label{subsec:strange}

Having presented BT's mechanisms, we now illustrate the heterogeneous download rate 
phenomenon and its consequences with two simple examples. Consider a swarm formed 
by a seed and 5 leechers. All peers, including the single seed, have identical 
upload capacity (64 kBps), but large (unconstrained) download capacity. The leechers 
download a file containing 1000 pieces (256 MB) and exit the swarm immediately after 
download completion. The seed never leaves the swarm.
This system was evaluated using an instrumented implementation of the BitTorrent
mainline 4.0.2 client (also used in \cite{legout_urvoy-keller_michiardi_2006}) running on
PlanetLab as well as a
detailed packet-level simulator of BT.
Both the PlanetLab experiments and the simulations use fully functional BT
clients that implement all BT control messages and policies, including peer selection strategies:
TFT, optimistic unchoke; and piece selection modes: random-first, rarest-first,
strict priority.

The simulation model was developed in the modeling tool Tangram-II
\cite{tangram2} (open source and publicly available software).
The model we developed is very detailed and faithfully implements the
protocol of the BitTorrent mainline 4.0.2 client, including all control 
messages and policies. In accordance with Tangram-II's modeling paradigm, 
entities that participate in the system are implemented as separate
objects that communicate by message passing. Thus, peers (leechers and seed) 
and tracker are represented by objects that can be fully parametrized (upload 
rate, file size, seed after download, etc). 

\begin{figure}[!t]
\begin{center}
\captionsetup[subfloat]{}
\subfloat[Arrival intervals: 10 min, 4 min, 4 min, 4 min.]{\label{fig:simul1}
\includegraphics[width=2.8in]{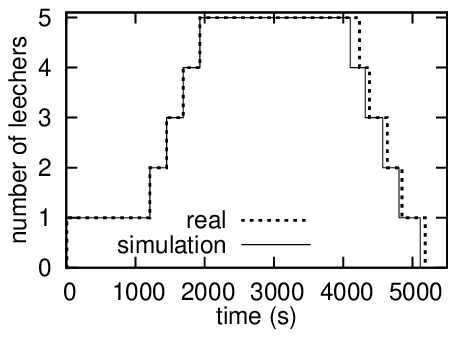}}
\subfloat[Arrival intervals: 4 min, 4 min, 4 min, 10 min.]{\label{fig:simul2}
\includegraphics[width=2.8in]{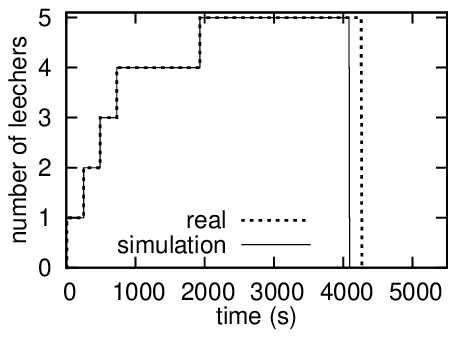}}
\caption{\label{fig:swarm-size} Evolution of the number of leechers in the swarm.}
\end{center}
\end{figure}

In the following simulations and experiments, leechers start to join the swarm only after the seed is
connected and they leave immediately after finishing the download.
The simulation/experiment ends when the last leecher leaves.
Figures \ref{fig:simul1} and \ref{fig:simul2} show the evolution of the swarm size as a function 
of time for both simulation and experimental results and two different leecher
arrival patterns. In Figure \ref{fig:simul1}, peers leave the swarm in the order
they arrived (i.e., FIFO) and have a relatively similar download time. Thus, the
download time is relatively indifferent to arrival order (with the exception of
the first peer).

Figure \ref{fig:simul2} shows the same metric just for different arrival 
times (in fact, the inter-arrival times of peers are also mostly preserved). 
Surprisingly, an unexpected behavior can be observed in the system dynamics:
despite the significant difference on arrival times, all five leechers completed
their respective download nearly at the same time.
The time inter departures is small comparing to the download time, which characterizes bursty departures. It means that peers that arrive later to the swarm have a smaller download
time. In fact, the fifth peer completed the download in about half the time of
the first leecher. Thus, the system is quite unfair with respect to the arrival
order of leechers, with late arrivals being significantly favored.
What is happening? Why does BT exhibit such
dynamics? We answer these questions in the next sections.

%
%
%

\section{Heterogeneity in homogeneous BT swarms}
\label{sec:problem}

\begin{figure}[!t]
\begin{center}
\captionsetup[subfloat]{margin=0.5pt}
\subfloat[Corresponding to Figure~\ref{fig:simul1}.]{\label{fig:simul1_downloaded}
\includegraphics[width=2.8in]{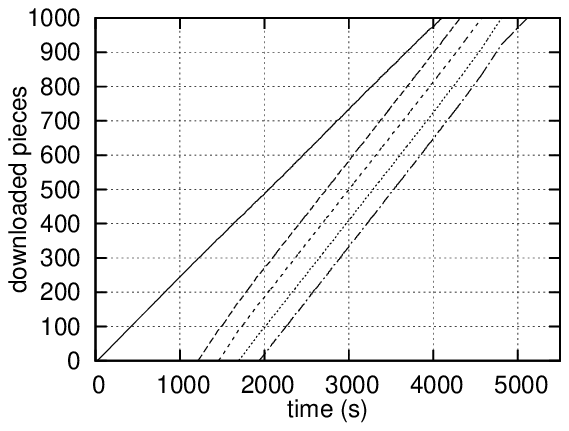}}
\subfloat[Corresponding to Figure~\ref{fig:simul2}.]{\label{fig:simul2_downloaded}
\includegraphics[width=2.8in]{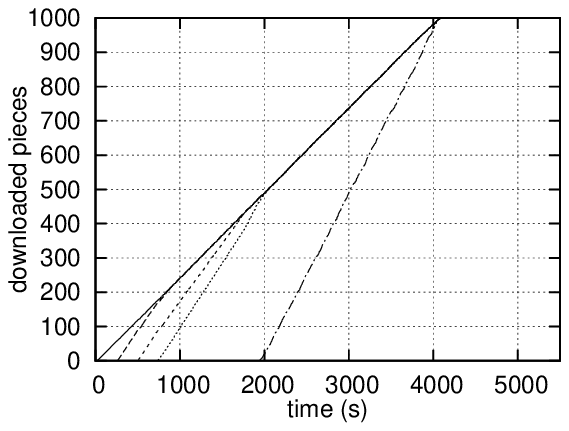}}
\caption{\label{fig:pieces-downloaded} Evolution of the number of downloaded pieces.}
\end{center}
\end{figure}

In order to understand the behavior exhibited by BT in Figures 
\ref{fig:simul1} and \ref{fig:simul2}, we will analyze the total number 
of pieces each leecher has downloaded over time.
Consider Figures~\ref{fig:simul1_downloaded} and \ref{fig:simul2_downloaded} where each
curve indicates
the total number of pieces downloaded by a given peer for the corresponding
scenario in Figures~\ref{fig:simul1} and \ref{fig:simul2}, respectively.
Note that the slope of each curve corresponds to respective leecher's
download rate.

We start by considering Figure~\ref{fig:simul1_downloaded}.
Despite the slope of the first leecher being 
smaller than that of the remaining peers, the curves never meet. In particular, 
a leecher finishes the download (and leaves the swarm) before the next 
leecher reaches the number of blocks it has. We also note that all other leechers 
have very similar slopes. In addition, we observe a peculiar behavior: the slope of 
the fifth leecher suddenly decreases when it becomes the single leecher in the 
system. 

The results illustrated in Figure~\ref{fig:simul2_downloaded} which 
correspond to the scenario considered in Figure~\ref{fig:simul2} show a 
very different behavior. Several
interesting observations can be drawn from this figure. The slope 
of the first peer is practically constant, remaining unchanged by the arrival 
of other peers. The slope of all other peers is larger than that of 
the first peer, meaning the curves may eventually meet. When two curves meet, the 
corresponding  leechers have the same number of blocks and possibly the 
same content (we will comment on this point in the following section). 
The figure also shows that a younger peer does not overcome the first peer, 
but instead the two maintain the same number of downloaded pieces, possibly 
with their contents synchronized. 
Finally, the slope of the second, third and fourth peer are rather similar.
However, the slope 
of the fifth peer is slightly larger than the others, meaning a higher download rate and 
consequently smaller download time. 

%
%
%

In summary, we make the following general observations:
\begin{itemize}
\item 
The first leecher downloads approximately at constant rate.
\item 
Subsequent leechers download at a faster rate than the first.
\item 
Once a leecher reaches the total number of pieces downloaded by the first 
leecher, their download rates are identical.
\item 
The greater is the number of leechers with the same number of pieces of the first 
leecher, the higher is the download rate of the other leechers.
\end{itemize}
All these observations are related to the dynamics of BT and will be 
discussed and explained in Section~\ref{sec:model} using a simple mathematical 
model. In the remainder of this section, we discuss the consequences of 
the observed phenomenon and illustrate that it happens even when peer arrival is 
random (i.e., Poisson process).

\subsection{Consequences of heterogeneity in homogeneous swarms}
\label{subsec:consequences}

Despite the homogeneous upload 
capacity of peers, the observations above lead to the following consequences:
\begin{itemize}
\item
{\bf Variability in download times.} Since peers can experience a 
consistently different download rate, their download times can also differ. 

\item
{\bf Unfairness with respect to peer arrival order.} Since peers download rates, 
and thus download times, may depend on their arrival order, the system is inherently 
unfair, potentially benefiting latecomers in a swarm. 

\item
{\bf Content synchronization.} Due to different download rates and BT's piece 
selection mechanisms (most notably rarest-first), leechers can synchronize on 
the number of pieces they have and, more strongly, on the content itself. This 
means that peers may end up with exactly the same content at some instant,
despite arriving at different points in time.

\item
{\bf Bursty departures.} A direct consequence of content synchronization is bursty 
departures. This means that peers tend to leave the swarm within a small
interval of time despite arriving at the swarm at relatively far apart instants.

\end{itemize}

Although the figures do not show the content synchronization
explicitly, since the first leecher is downloading the file at the same rate at
which the seed pushes new pieces into the swarm (seed upload capacity), whenever a leecher reaches the
same number of pieces than it, they have exactly the same content.

Of course, the prevalence of the phenomenon and its consequences depend directly on 
the parameters of the swarm. In particular, the arrival times of peers is certainly 
the most determinant. However, parameters like upload capacity of seed and leechers 
and number of pieces are also fundamentally important. Intuitively, a file with a 
larger number of pieces or a seed with a lower upload capacity increase the
probability that the consequences above occur. In fact, for any arrival 
order of a small set of peers, one can always find system parameters for which
this behavior and its consequences occur.

\subsection{Heterogeneity under Poisson arrivals}
\label{sec:poisson}

The behavior above does not require deterministic 
arrivals or any crafted leecher arrival pattern. It arises even 
when arrival patterns are random. In this section we characterize the
consequences of the heterogeneous download rates 
phenomenon under Poisson arrivals.

We conducted a large amount of evaluations using detailed packet-level
simulations.
In particular, we consider a BT swarm where 
a single seed is present at all times, while leechers arrive according to 
a Poisson process and depart the swarm as soon as their download is completed. 
In the evaluation that follows, all leechers have the same upload capacity of 
64 kBps (and very large download capacities) and download a file with 
1000 pieces. The upload capacity of the seed ($c_s$) varies between 
48 kBps, 64 kBps, and 96 kBps, and the leecher arrival rate ($\lambda$) is
1/1000 s. 
These scenarios generate a swarm that has a time average size of 3.7, 3.4 and 3.0 leechers, 
respectively. 

\begin{figure}[!t]
\centering
\includegraphics[width=2.8in]{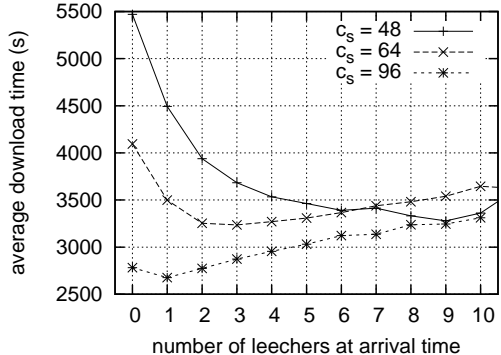}
\caption{Average download time as a function of arrival order in a busy period.}
\label{fig:xx-64_1000_1000_relative-order}
\end{figure}

We start by characterizing the variability in the download times and the 
unfairness with respect to leecher arrival order. Figure \ref{fig:xx-64_1000_1000_relative-order} 
illustrates the average download time for a peer as a function of the number
of leechers in the swarm at its arrival time. Thus, if a peer joins the swarm
when $i$ leechers are present, it is mapped to index $i$. 
The different curves correspond to different upload capacities of the seed. The 
results clearly indicate that the download time depends on leecher arrival 
order. In particular, for the case $c_s = 64$ kBps, the average download 
time tends to decrease with increasing arrival order, and so the first arrival 
has the largest average download time. Moreover, the download time 
differences are also significant, and can reach up to 30\% (e.g., difference 
between first and fourth arrival).

Figure~\ref{fig:xx-64_1000_1000_relative-order} also indicates that variability in download times strongly 
depends on the seed upload capacity. In particular, a fast seed yields the reverse 
effect: leechers' download times tend to {\it increase} with arrival order. Intuitively, when a 
slow seed is present, late arrivals to a busy period obtain large download rates from other 
leechers, thus exhibiting a lower download time. However, when a fast seed is present, 
the first leecher has the larger upload capacity of the seed until the second arrival, 
thus exhibiting a lower download time. The results also illustrate second order 
effects. For instance, a very late arrival can have an average download time slightly larger 
(or smaller) than a late arrival (e.g., the sixth leecher arrival has longer
download time than fourth for $c_s = 64$ kBps). 
Intuitively, this occurs because a very late arrival is likely to be alone in the busy 
period, having to resort to the seed for finishing the download. Since the upload capacity 
of the seed can be smaller (larger) than the aggregate download rate it receives from 
other leechers, its download time can increase (decrease). This behavior and its
consequences will be explained 
and captured by the mathematical model presented in the next section.

\begin{figure}[!t]
\begin{center}
\captionsetup[subfloat]{margin=0.1pt}
\subfloat[Empirical CCDF of peer inter-departure time conditioned on a busy period.]{\label{fig:xx-64_1000_1000_departure-ccdf}
\includegraphics[width=2.8in]{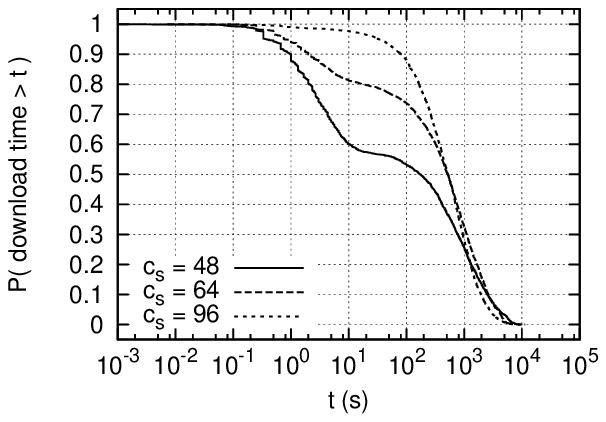}}
\subfloat[Empirical CCDF of the download time.]{\label{fig:download_times-cs}
\includegraphics[width=2.8in]{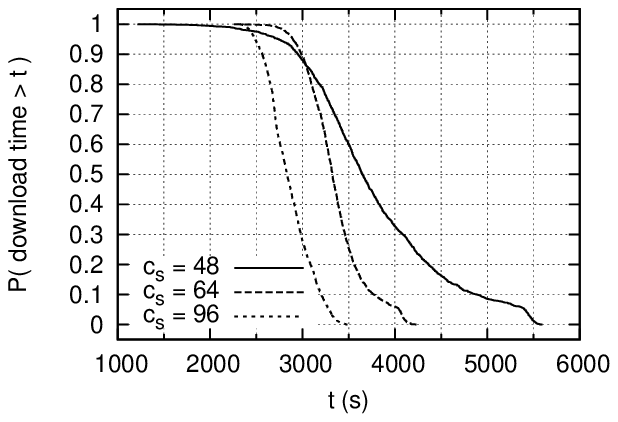}}
\caption{\label{fig:empirical_ccdfs}
Characterization of consequences
    of heterogeneous download rates for different values of seed capacity.}
\end{center}
\end{figure}

We now characterize the burstiness in the leecher departure process.
Figure~\ref{fig:xx-64_1000_1000_departure-ccdf} 
shows the empirical CCDF (Complementary Cumulative Distribution Function) of the leecher 
inter-departure times conditioned on a busy period (i.e., not including the inter-departure 
time between the last leecher in a busy period and the first leecher of the next). Note 
that the peer inter-arrival times follow an exponential distribution with rate 1/1000. However, 
the results indicate a very distinct departure process. In particular, many peers tend to 
leave the swarm at roughly the same time: up to 30\% of peers leave the swarm within a 
couple of seconds from each other when $c_s = 64$ kBps. Moreover, the departure process 
also exhibits high variability and some peers take as much as ten times more to
leave the system after a departure than the average 
(when $c_s = 64$ kBps). The figure also clearly shows that this observation strongly depends on 
the seed upload capacity, and is more pronounced when the seed is slow. Intuitively, a 
slower seed increases the average download time, thus increasing the chances that 
leechers synchronize their content during the download and depart almost at the same time.
Finally, we also note that a fast seed yields a much less bursty departure process, although 
still favoring short inter-departure times.

\begin{table}[h]
\centering
\caption{Average number of leechers and average number of synchronized leechers
conditioned on intervals where the number of leechers is greater than 1.}
\label{tab:synch}
\begin{tabular}{c|c|c}
$c_{s}$ & cond. avg. number & cond. avg. number\\
(kBps) & of leechers  & of synch. leechers\\
\hline
48 & 4.45 & 2.40 \\
64 & 3.86 & 1.44 \\
96 & 3.57 & 0.87 \\
\end{tabular}
\end{table}
One consequence of the heterogeneous download rates that is closely related to
the bursty departures is content synchronization. Here we refer to as
synchronized, leechers that are not interested in more than 50 pieces
(5\% of the file) of any other. In this context, we compare the
average number of leechers in the system and the average number of those which
are synchronized. These metrics are conditioned on time intervals where the
number of leechers is greater than 1, because synchronization is not
defined otherwise. Table \ref{tab:synch} shows the results of our simulations.
The conditional average number of synchronized leechers corresponds to $53.9\%$,
$37.3\%$ and $24.4\%$ for $c_s$ equal to $48$, $64$ and $96$ kBps
respectively. While the synchronization is
less pronounced when the seed capacity is high, it is very significant when $c_s
\leq c_l$.

It is possible to have different download times even when all peers
that are simultaneously in the swarm have the same instantaneous download rate.
Since peers join the system at different times, they observe the swarm in
different sequences of states, in some of which there is more bandwidth
available. Those peers will have smaller download times.
Nevertheless, as we discussed in Section~
\ref{subsec:consequences}, heterogeneous download rates also contribute to the
variability in the download times. Figure~\ref{fig:download_times-cs} shows the
empirical CCDF of the leecher download time for different values of seed
capacity ($c_s$). While the maximum download
time is $45.5\%$ and $52.8\%$ higher than the minimum respectively for $c_s$ equal to $96$
and $64$ kBps, it is $218.7\%$ higher for $c_s = 48$ kBps. Surprisingly, the
minimum download time is the smallest when the seed capacity is the lowest
(i.e., 48 kBps). This is because leechers synchronize with high
probability under these circumstances and, as we will see in
Section~\ref{sec:validation}, non-synchronized
leechers receive at very high download rates in the presence of many
synchronized ones.

We observe that the seed capacity plays an important role on the occurrence of the
described consequences under unpopular swarms. In the following, we characterize the
impact of another important aspect on these consequences, namely the content
popularity, which can be captured through leecher arrival rate. For this purpose, we conducted simulations where
the seed and the leechers have the same upload capacity of 64 kBps and
average inter-arrival time (i.e, $1/\lambda$) varying between
500, 1000, 1500, 2000 and 2500 s.

\begin{figure}[!t]
\begin{center}
\captionsetup[subfloat]{margin=0.1pt}
\subfloat[Box plot of download time of leechers.]{\label{fig:64-64_xxxx_1000_box-plot}
\includegraphics[width=2.8in]{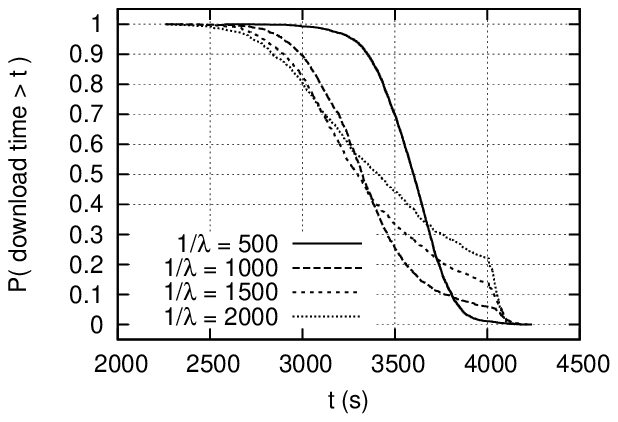}}
\subfloat[Conditional average number of leechers and conditional average number of synchronized leechers.]{\label{fig:xx-64_xxxx_1000_synch}
\includegraphics[width=2.8in]{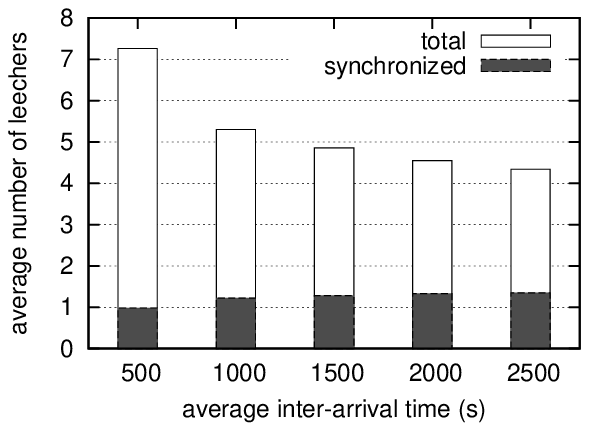}}
\caption{\label{fig:interarrival_results} Characterization of consequences
    of heterogeneous download rates for different values of average inter-arrival time.}
\end{center}
\end{figure}

We consider the influence of the average inter-arrival time of leechers on the download times, 
independently of arrival order. Figure~\ref{fig:64-64_xxxx_1000_box-plot} shows
the empirical CCDF 
of the download times 
of peers as a function of the average inter-peer arrival time (i.e., the inverse
of arrival rate), for $c_s = 64$ kBps. Note that there are sharp drops for $t >
4000$ which correspond to leechers whose average download rate is approximately
equal to $c_s$. These sharp drops are more pronounced when the inter-arrival
time is large. In addition, as the inter-arrival time grows, the
10th-percentile decreases and the 90th-percentile increases, indicating
that the download times become less concentrated around the average.
However, the variability between minimum and maximum
download time does not diminish with the inter-arrival time.

Figure~\ref{fig:xx-64_xxxx_1000_synch} illustrates the intensity of content
synchronization for different arrival rates. It shows the
average number of leechers in the system and the average number of those which
are synchronized. We observe that,
the number of synchronized leechers
remains practically the same as we increase the inter-peer arrival time,
indicating that a larger fraction of peers have similar content when popularity
decreases.

As with content synchronization, the fraction of bursty departures is also
strongly dependent on the leecher arrival rate. While approximately $5\%$ of the
intervals between departures are smaller than 10 seconds for an arrival rate
$\lambda = 1/500$, more than $30\%$ of intervals are smaller than 10 for
$\lambda = 1/2500$. On the other hand, the unfairness
with respect to the arrival order in a busy period is almost insensitive to the leecher
arrival rate (considering $1/2500 \leq \lambda \leq 1/500$).

\subsection{Real experimental evaluation}

The results shown above were all obtained through simulations and we now 
present results from prototype-based experiments deployed in more realistic 
scenarios. The real experiments were performed in the Internet using machines
from Planetlab\cite{planetlab} and running an instrumented version of a BT
client\cite{legout_urvoy-keller_michiardi_2006}. Although a large number of
experiments were conducted, we 
report only on a limited set of these results due to space constraints. The goal here 
is to validate the phenomenon of heterogeneity in homogeneous BT swarms and its 
consequences in real BT application running over the Internet.  

In the experiments, the PlanetLAB machines were selected using a quick and
simple performance test. Before starting every experiment, a controller
dispatches a command via ssh for a set of few hundred machines randomly
chosen from the complete list of all PlanetLAB machines. The command line
basically makes the machines to download and install all the necessary files
(including BT client and scripts) to execute locally the experiment. The set of
machines that had the best performance downloading and installing the files was
used in the experiments. This performance test was enough to avoid using
machines overloaded or connected through congested links.

We consider only private swarms in the experiment, in the sense that only peers 
controlled by the experiment can connect to the swarm for uploading and downloading 
content. Each private swarm consists of a single file of size $S$~MB which is owned by a 
single seed that is always available and has upload capacity of $c_s$. Leechers
interested in  downloading the content arrive to the swarm according to a
Poisson process with rate $\lambda$. 
All leechers that arrive to the swarm are homogeneous and have upload capacity equal to $c_l$.
The maximum upload capacities used in the experiments are defined as parameters
of any BT client (including the one we use). Note that those upload capacity values used for the
experiments were far below the limit imposed for each slice (user) in PlanetLAB.
Each experiment run is executed for $t=5,000$~seconds and leave the swarm once
the download is completed.
 
\begin{figure}[!t]
\begin{center}
\captionsetup[subfloat]{margin=0.1pt}
\subfloat[Evolution of swarm size.]{\label{fig:swarm_size_experiment}
\includegraphics[width=2.6in]{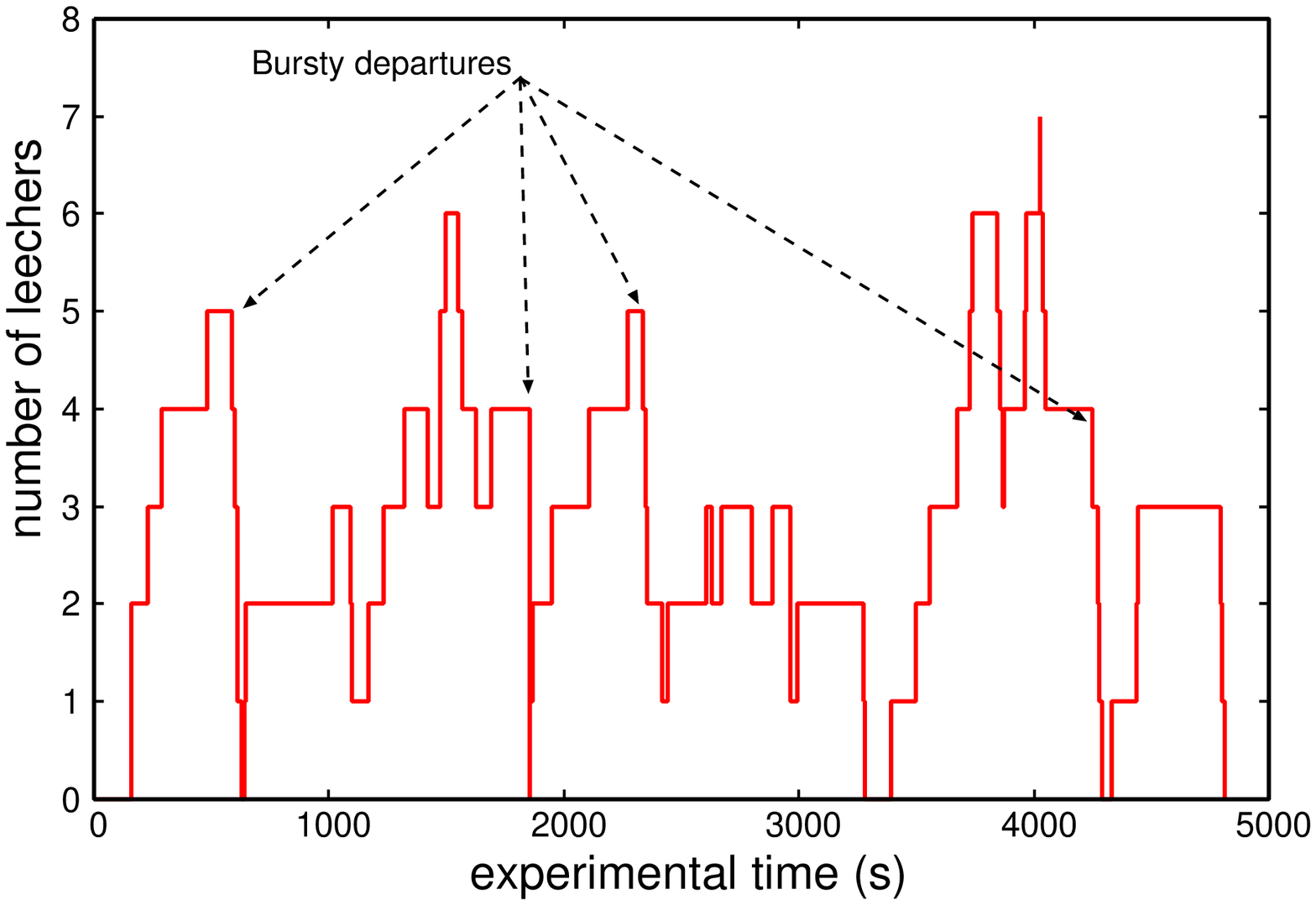}}
\subfloat[Leechers' arrivals and
departures.]{\label{fig:dynamic}
\includegraphics[width=2.6in]{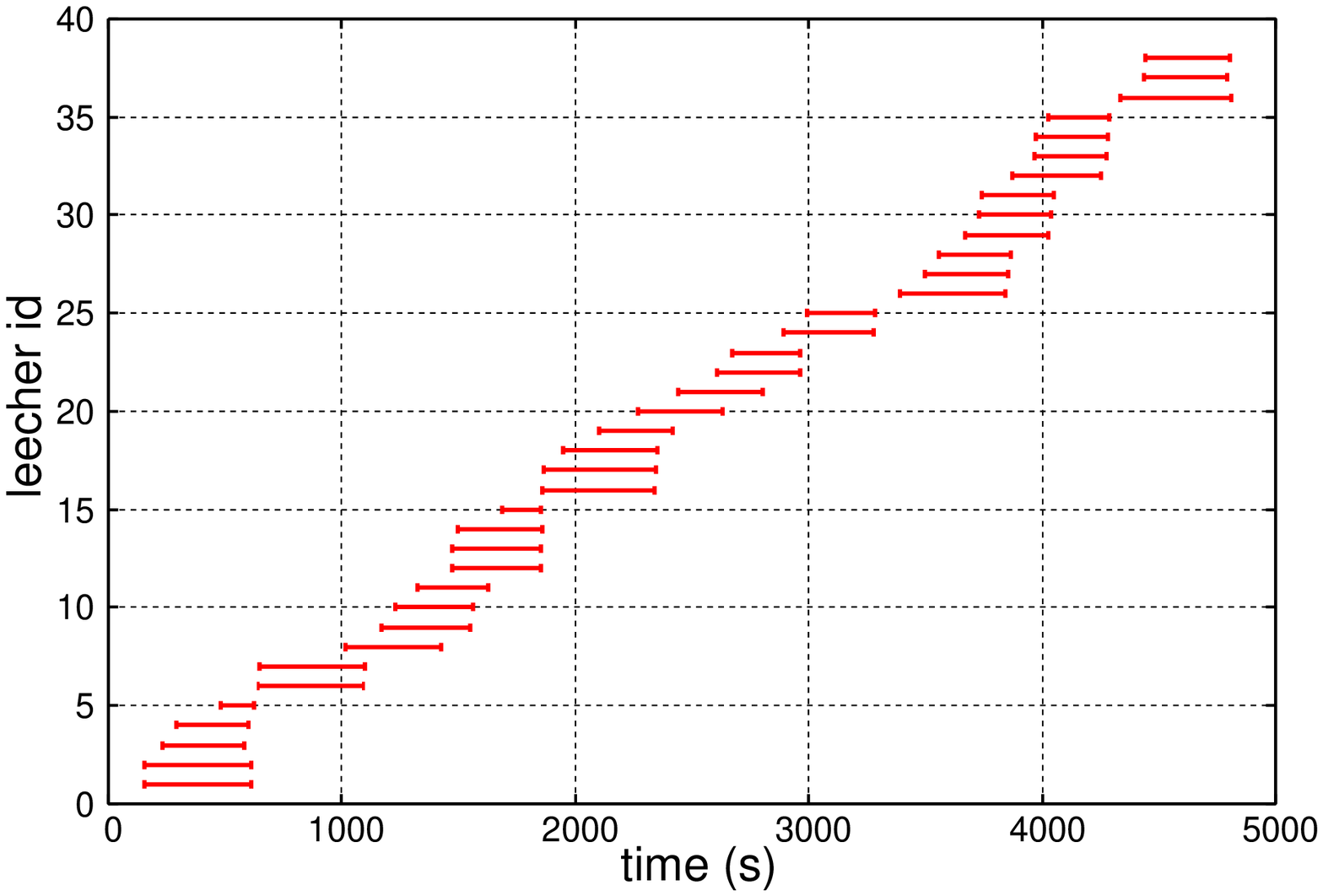}}\\
\subfloat[Zoom-in of the first busy period.]{\label{fig:swarm_size_zoom}
\includegraphics[width=2.6in]{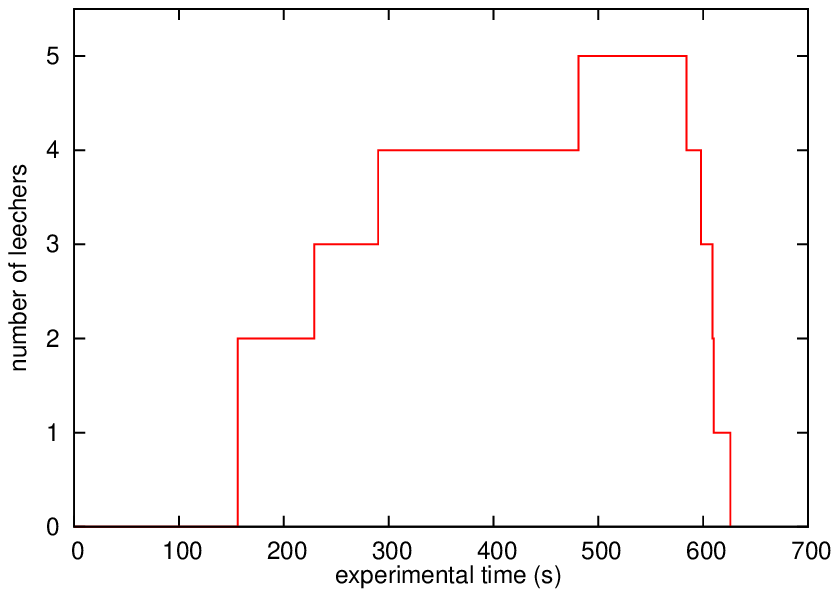}}
\caption{\label{fig:} Swarm dynamics in real experiments.}
\end{center}
\end{figure}

We start by analyzing the evolution of the swarm size for an unpopular swarm.
Figure~\ref{fig:swarm_size_experiment} shows the number of leechers in the swarm
over time for the duration of the experiment, with parameters $\lambda=1/125$
peers/s, $S=20$~MB, and $c_s=c_l=50$~kBps. We can observe several occurrences of
bursty departures, even if leechers arrive according to a Poisson process. As
previously discussed, bursty departures are consequence of content
synchronization among the leechers in the swarm.
 
Using the same experiment as above, we investigate the impact of the leechers' arrival order 
on their download times. Figure~\ref{fig:dynamic} illustrates the dynamics of the swarm, where 
each horizontal line corresponds to the lifetime of a leecher in the swarm, starting when the 
peer arrives and ending when it departs the swarm. Note that peers exhibit significantly 
different download time (which corresponds to their lifetime in the system). In particular, 
in many cases leechers arrive at different time instants but depart in the same burst. 
For instance, the fifth leecher to arrive to the swarm departs in a burst
almost together with all 
four prior arrivals (see Figure~\ref{fig:swarm_size_zoom} for a zoom-in of the
first busy period). Thus, the fifth leecher has a much smaller download completion time, when 
compared to the first leecher. Similar behavior occurs between the fifteenth leecher and the 
three leechers that arrived immediately before. Besides illustrating the variability of the 
download times, this observation also indicates the unfairness with respect to leecher arrival 
order. In particular, late arrivals to a busy period tend to have smaller download times. 

\begin{figure}[!t]
\begin{center}
\captionsetup[subfloat]{margin=0.1pt}
\subfloat[$c_s=50$~kBps and $c_l=50$~kBps.]{\label{fig:download_ccdf_50}
\includegraphics[width=2.8in]{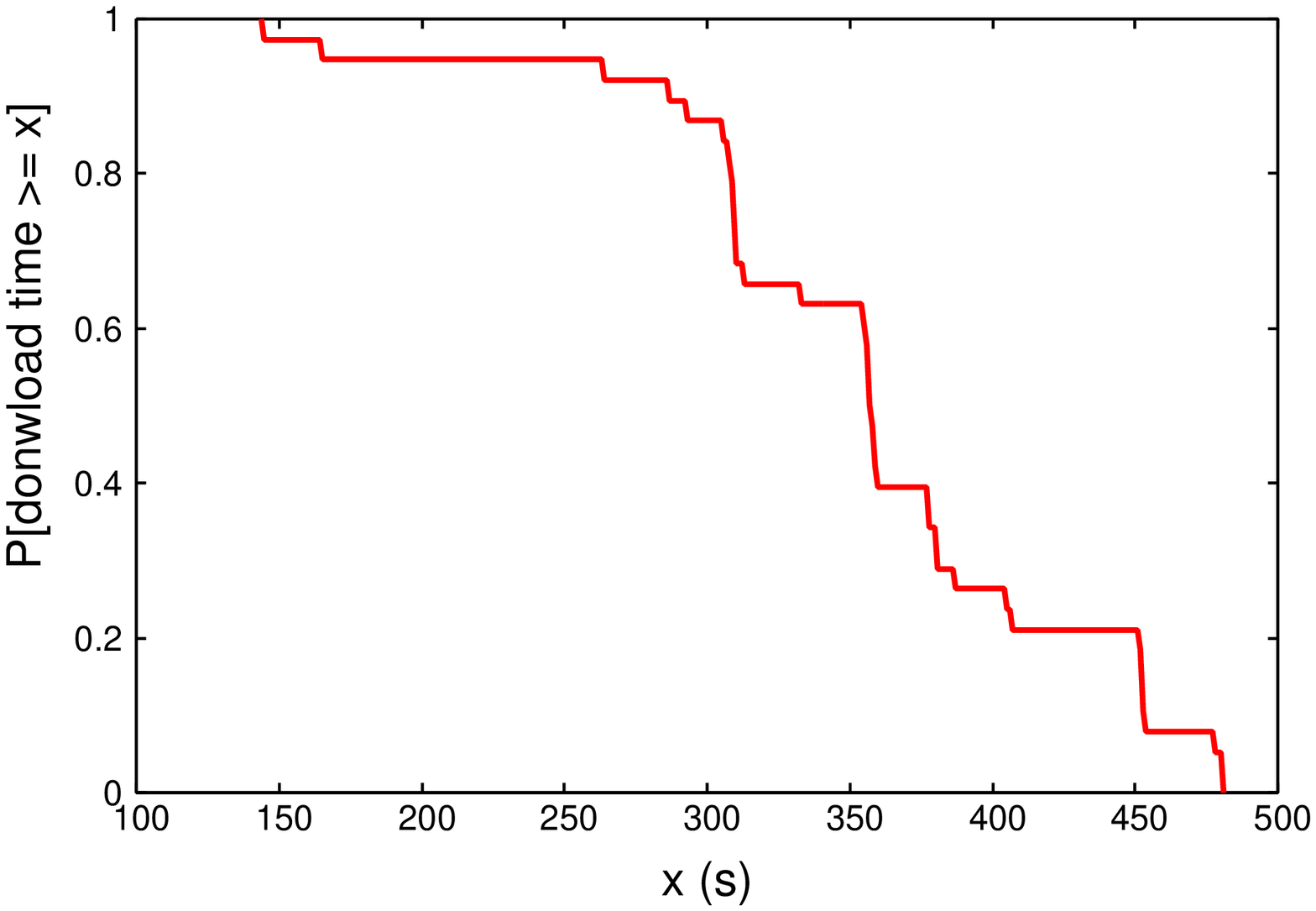}}
\subfloat[$c_s=60$~kBps and $c_l=50$~kBps.]{\label{fig:download_ccdf_60}
\includegraphics[width=2.8in]{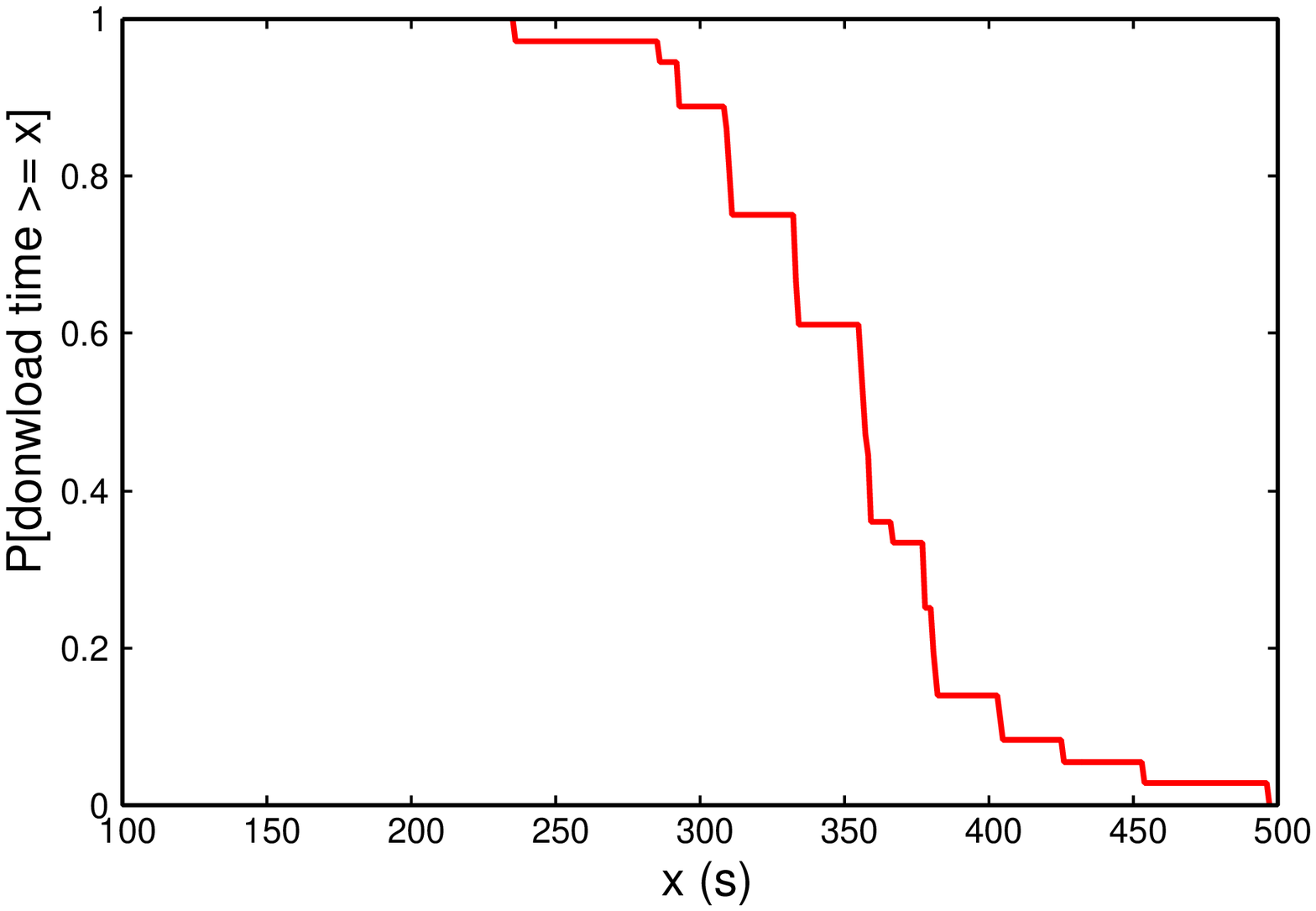}}
\caption{\label{fig:download_ccdf} CCDF of download time from real experiments.}
\end{center}
\end{figure}

We now focus on the distribution of the leechers' download times to illustrate their 
relative high variability. Figures~\ref{fig:download_ccdf_50} and
\ref{fig:download_ccdf_60} show the complementary cumulative distribution function (CCDF) of download times computed for two experiments with 
distinct upload capacities for the seed ($c_s = 50$~kBps and $c_s = 60$~kBps, respectively, 
with all other parameters the same). In both results, download times exhibit a high variance, 
as shown in the figures. In the case $c_s=50$~kBps (Figure~\ref{fig:download_ccdf_50}), the 
minimum and maximum values are 145 and 480 seconds, respectively, with the maximum being more 
than three times the minimum. When the upload capacity of the seed is higher
than that of the 
leechers, Figure~\ref{fig:download_ccdf_60} shows that the variance in download times 
decreases, as expected, since the system capacity is increased. Finally, we note 
several discontinuities (i.e., sharp drops) in both CCDF curves which are caused 
by sets of leechers that have approximately the same download time.

\section{Model}
\label{sec:model}

We develop a simple model attaining to understand the origin of the heterogeneous
download times and its consequences. Our model obtains an approximation to the 
average upload and download rates observed by each leecher on different time 
intervals for unpopular swarms.


Consider a homogeneous swarm of some unpopular content with a single seed to which
leechers arrive sequentially and depart as soon as they complete their download,
such as the one illustrated in Figure~\ref{fig:simul1}. By unpopular content we
imply a swarm with an arrival rate that is small enough such that there is never 
too many peers in the swarm. In particular, our modeling framework assumes that 
the maximum number of upload connections of peers is always larger than (or
equal to) the swarm 
size. In such scenario, Tit-for-Tat (TFT) and optimistic unchoke algorithms have 
no effect, since all peers upload to one another. Thus, such mechanisms are not 
present in our model. However, note that rarest-first mechanism continues to operate 
since is not affected by this assumption and is therefore captured by our model. 

In the described scenario, bursty departures can only happen if younger leechers
obtain roughly the same number of pieces as older ones, and leave the swarm 
at about the same instant. This in turn implies that younger leechers 
must have higher download rates than older ones, at least for some
periods of time. Why is that? At a given moment, an older leecher $i$ may 
have all pieces owned by a younger leecher $j$. Thus, leecher's $j$ uplink 
capacity will be used to serve other leechers until $j$ receives a piece 
that $i$ does not have. During this period of time, $j$ simply cannot 
serve $i$, even if it has no other leecher to serve. Therefore, the 
sets of pieces owned by each leecher are the root causes for
heterogeneous download rates and must be considered. 

\begin{figure}[h]
\centering
\includegraphics[width=2in]{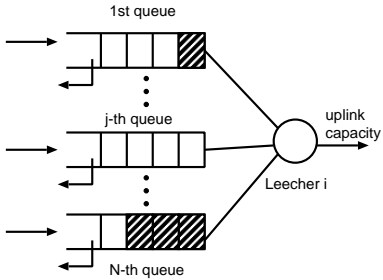}
\caption{Leecher $i$ can be represented as server with multiple queues, one for each
neighbor, containing pieces that are interesting to them.}
\label{fig:multiple-queue}
\end{figure}

In order to capture the observation above, each peer, either a seed or a
leecher, is represented by a queueing system with multiple queues (see
Figure~\ref{fig:multiple-queue}), one for each neighbor, under a
processor sharing discipline. Queue $j$ of peer $i$ contains the pieces
interesting to peer $j$ (i.e., all pieces that $i$ has that $j$ has not). When
peer $j$ downloads one of these pieces, from $i$ for instance, this piece is
removed from the $j$-th queue of $i$, and from the $j$-th queues of other peers where the
piece was present. On the other hand, whenever a peer downloads a piece that
other neighbors are interested in, this piece is placed in the queues
corresponding to those neighbors, increasing their queues sizes. Finally, the
queues of the seed always have all pieces that are needed by the leechers. As a
leecher downloads pieces from the seed and other leechers, this queue decreases,
eventually becoming empty when the leecher downloads the entire content
and departs the swarm. We note that the order at which these pieces are
served from these queues depend on the piece selection policy. 

Let $c_s$ and $c_l$ be the seed and leechers' uplink capacities, respectively.
Assume that the leechers' downlink capacities are much larger than $c_s$ and
$c_l$. Let $N(t)$ be the number of leechers in the system at time $t$. Since 
the seed always has interesting pieces to every leecher, all the $N(t)$ queues 
in the seed are backlogged. Thus, all queues will be served at rate $c_s/N(t)$. 
Note that, since the swarm is unpopular, we assume the swarm size is small 
enough such that every leecher is neighbor of every other peer (including the
seed) and can serve all of them simultaneously. 

A leecher may not have interesting pieces to some of its neighbors at time $t$.
Let a leecher be identified by its arrival order, thus leecher $i$ is the $i$-th
leecher to join the swarm. Also let $n_i(t) \leq N(t)-1$ be the number of
leechers interested in pieces owned by $i$. The instantaneous upload rate 
from $i$ to any of these leechers is $c_l/n_i(t)$.

Whether a leecher has or has not pieces interesting to another
depends on the leechers' respective {\it bitmaps}, i.e. the current 
subsets of pieces owned by a leecher at time $t$. The set of bitmaps of all leechers 
would precisely determine the exact pieces in each queue. However, 
the dynamics of the bitmaps are intricated and to keep track of them would
be unnecessarily complicated for modeling the phenomenon we are interested in.
Instead, we consider the number of pieces owned by each leecher $i$, $b_i(t)$,
    and infer whether a leecher has interesting pieces to other leechers. 

For the sake of simplicity, let $b_i(t) = b_i$ and $N(t) = N$. 
Two remarks can be made with respect to $b_i$ and the interest relationship among 
leechers:
\begin{remark}[]
\label{more_pieces}
If $b_i > b_j$, then $i$ has at least $b_i - b_j$ interesting pieces to $j$. 
\end{remark}
\begin{remark}[]
\label{less_pieces}
If $0 < b_i \leq b_j$, it is impossible to determine whether $i$ has or 
has not interesting pieces to $j$ without further information.
\end{remark}

In the following, we will use these two remarks to derive a simple model to 
capture the upload and download rates between peers. With respect to 
Remark \ref{less_pieces}, we will assume that no further information is available, 
and hence the piece interest relationship among peers will be ignored in this 
case. Nevertheless, we will see that a peer with less pieces than other can
still upload pieces to the latter.

\subsection{A simple fluid model}

We assume that content is fluid, or equivalently, that pieces can be
subdivided in infinitely many parts that can be exchanged (uploaded 
and downloaded) continuously. 


To simplify the explanation, assume that $b_1 > b_2 >
\dots b_N$, i.e. an older leecher has strictly more pieces than a younger 
one. We will relax this assumption later on this section, allowing the model to
represent swarms where two peers arrived at the same time, or more
generally, where some leechers have the same number of pieces. 
We now make the following assumptions:

\begin{itemize}
\item Even if leecher $i$ has joined the swarm after $j$, i.e.
$i > j$, $i$ can still upload pieces to $j$ as long as $i$ downloads 
pieces from any peer $k$ that has more pieces than $j$, i.e. $k < j$. Thus, 
younger peers can upload to older peers. 

\item Every piece downloaded from the seed by a leecher is 
immediately interesting to all other leechers, independently of their 
arrival time. The rarest-first piece selection policy provides support 
for this assumption.
Figure~\ref{fig:example-interesting-pieces} depicts the idea that a younger peer
can upload pieces to an older one. In this scenario, peer 4 can upload to peer 2,
since it is downloading pieces interesting to peer 2 from the seed and from peer 1. 

\begin{figure}[!t]
\centering
\includegraphics[width=2.0in]{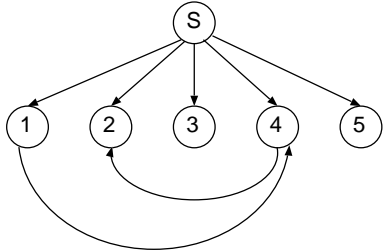}
\caption{Peer 4 can upload pieces to peer 2, since it is downloading pieces
interesting to the latter from the seed and from peer 1.}
\label{fig:example-interesting-pieces}
\end{figure}


%

\item Since the seed's upload capacity is $c_s$, each leecher downloads from 
it at rate $c_s/N$. Now let $g_{ij}$ be the rate at which peer $i$
could potentially upload data to peer $j$ provided that there is no capacity 
constraints (i.e. independently of upload and download capacities of 
peers $i$ and $j$, respectively).
If a leecher $i$ is older than $j$, $i$ has interesting pieces to $j$.
Therefore, from the perspective of the multiple queueing system, queue $j$ 
in leecher $i$ is backlogged and $g_{ij} = \infty$. 
On the other hand, if $i$ is younger than $j$, the rate
$g_{ij}$ is given by the rate at which $i$ downloads interesting 
pieces to $j$.
\end{itemize}

We draw the reader's attention to the first two assumptions. They account for
the upload rate that a younger leecher can sustain to an older leecher, even
though we cannot say that the former has interesting pieces to the latter just
from the number of pieces they own. 

From these assumptions, we can conclude that the rate $g_{ij}$ at 
which a peer $i$ uploads interesting pieces to an older peer $j$ 
is equal to the rate at which peers older than $j$ upload to 
$i$ plus the rate at which $i$ downloads from the 
seed. We thus have:
\begin{subnumcases}{\label{eq:bandwidth-requirements} g_{ij}=}
 \infty,     & if $i < j$\\
  \frac{c_s}{N} + \sum_{k<j}u_{ki},  & if $i > j$ \label{eq:bandwidth-requirement}
\end{subnumcases}
where $u_{ki}$ is the rate at which leecher $k$ uploads to $i$.

For instance, in Figure~\ref{fig:example-interesting-pieces}, $g_{4,2}$
is the sum of the rates at which peer 2 downloads from the seed ($c_s/5$) and
from peer 1 ($u_{1,4}$). Hence, $g_{4,2} = c_s/5 + u_{1,4}$. Again we see that
the proposed model accounts for the fact that a younger leecher can upload 
pieces to the older ones. In a real swarm however, peer 4 may upload to peer 2
pieces downloaded from younger leechers as well, such as peer 5. Although the
pieces that peer 5 downloads from the seed are immediately interesting to both
peers 2 and 4, they will not start and finish downloading this piece from peer 5
at the same time. Thus, leecher 4 may finish first the download of such a piece and 
then help serve the remaining sub-pieces to peer 2, violating our assumption. Intuitively
however, the contribution of peer 4 in uploading this piece to peer 2 is small, 
since peer 4 must fully finish the download before it can start uploading, by which 
time peer 2 will have downloaded most of the piece from peer 5. 
Thus, we claim that such effects are negligible and can be ignored since the model 
is accurate when compared to simulations and experiments, as discussed in 
Section \ref{sec:validation}.


We now make an important observation concerning Equation~(\ref{eq:bandwidth-requirement}). 
Consider leecher $i$ and some other leecher $j$. The older $j$ is with respect to 
$i$ the smaller is the rate at which $i$ can upload to $j$, that is, the smaller is 
$g_{ij}$. If $j$ is younger than $i$, then $g_{ij} = \infty$. This observation implies 
that $g_{i1} \leq g_{i2} \leq \dots \leq g_{iN}$.

\begin{figure}[!t]
\centering
\includegraphics[width=6.0in]{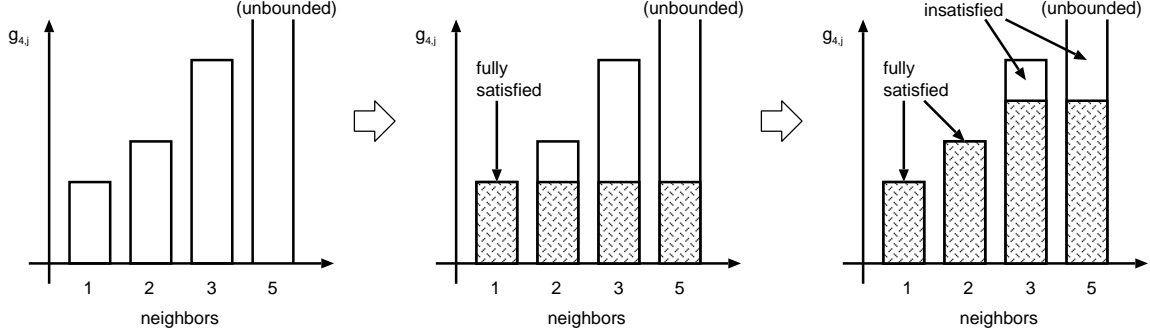}
\caption{The upload bandwidth allocation of leecher $i$ follows a progressive
filling algorithm.}
\label{fig:water-filling}
\end{figure}

In addition, note that $g_{ij}>0$ for all $i,j$. As we consider a
small swarm, all peers upload to one another.
Since the upload capacity of peers is finite, we must now determine 
how the capacity of a given peer $i$ will be divided to serve each of the $N-1$
other leechers. 
In particular, recall that $u_{ij}$ is the upload rate from peer $i$ to peer $j$ and note 
that $\sum_k u_{ik} \leq c_l$, where $c_l$ is the upload capacity of a leecher. To determine 
$u_{ij}$ given the values of $g_{ij}$, where $1 \leq j \leq N$,
we use a bandwidth allocation mechanism that
follows a progressive filling algorithm. This mechanism determines the outcome of the processor
sharing discipline. Figure \ref{fig:water-filling} illustrates the progressive
filling algorithm for the example presented in
Figure~\ref{fig:example-interesting-pieces}.
Roughly, infinitesimal amounts of bandwidth are allocated to each neighbor until
(1) the leecher's capacity is completely allocated or (2) a leecher $j$ is
satisfied with respect to the $g_{ij}$ constraint. In the former case, the
algorithm stops.  In the latter, it continues to distribute the remaining
capacity among the non-satisfied leechers until one of the two conditions occurs
again.

Due to the fact that $g_{i1} \leq g_{i2} \leq \dots \leq g_{iN}$,
the final bandwidth allocation for leecher $i$ can be efficiently 
obtained by computing the following equation in the order $j=1,\dots,N$:
\begin{equation}
\label{eq:model_equations}
u_{ij} = \min\Bigg( g_{ij},
    \frac{ c_l - \sum_{k<j}u_{ik}}
         { N-1-|\{k|k<j,k\neq i\}| }\Bigg)
\end{equation}
where $|A|$ is the cardinality of a set $A$.
Now recall from Equation~(\ref{eq:bandwidth-requirement}) that $g_{ij}$
depends on $u_{1,i},u_{2,i},\dots,u_{j-1,i}$, for $i > j$. Therefore, by calculating
$u_{ij}$ in the order $i=1,\dots,N$, we assure that every variable in
Equation~(\ref{eq:model_equations}) has been previously computed.

\begin{figure}[!t]
\centering
\includegraphics[width=1.7in]{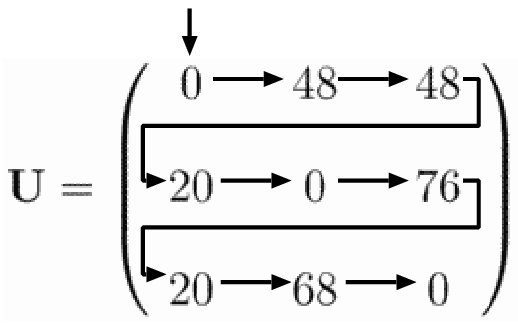}
\caption{Example of matrix $\mathbf{U} = (u_{ij})$ showing the right order of
    calculation.}
\label{fig:matrix}
\end{figure}

As an example, consider the calculation of the matrix $\mathbf{U} = (u_{ij})$, which 
determines upload rates between peers at a given moment, for a small swarm containing 
a single seed and $N = 3$ leechers. Let their upload capacities be equal 
to $c_s=60$ kBps and $c_l=96$ kBps, respectively, and assume $b_1 > b_2 > b_3$.
Matrix $\mathbf{U}$ and the order of computation of its elements are 
depicted in Figure~\ref{fig:matrix}. The download rate $d_i$ for peer $i$ is
simply $c_s/N$ plus the sum of the elements in column $i$:
\begin{equation}
\label{eq:download}
d_i = \frac{c_s}{N} + \sum_{j=1}^{N} u_{ji}.
\end{equation}
Hence,
\begin{eqnarray}
d_1 & = & 60/3 + 0 + 20 + 20 = 60 \label{d1} \\
d_2 & = & 60/3 + 48 + 0 + 68 = 136 \label{d2} \\
d_3 & = & 60/3 + 48 + 76 + 0 = 144 \label{d3}
\end{eqnarray}

Equations~(\ref{d1})-(\ref{d3}) corroborate the idea that homogeneous 
peers can exhibit heterogeneous download rates which depend on the number of 
pieces owned by each leecher. Moreover, younger leechers tend to 
have a higher download rate, as they obtain a higher upload rate from other 
leechers. This is the opposite of what happens in large swarms, where the older
leechers usually manage to keep the TFT for longer periods, hence achieving higher
download rates.

Eventually the number of pieces owned by a leecher may reach the number of 
pieces owned by an older one. In particular, this is bound to occur since 
younger leechers tend to have a higher download rate. In this case, these 
two leechers will no longer have pieces interesting to each other. Thus, 
Equations (\ref{eq:bandwidth-requirements}) and (\ref{eq:model_equations}) 
must be rewritten as functions of $b_i,\forall i$:
\begin{subnumcases}{\label{eq:bandwidth-requirements2} g_{ij}=}
 \infty,     & if $b_i > b_j$\\
   \frac{c_s}{N} + \sum_{b_k>b_j}u_{ki}, & if $b_i \leq b_j$
  \label{eq:bandwidth-requirement2}
\end{subnumcases}
\begin{equation}
\label{eq:model_equations2}
u_{ij} = \min\Bigg( g_{ij},
    \frac{ c_l - \sum_{k|b_k>b_j}u_{ik}}
         { N-1-|\{k|b_k>b_j,k\neq i\}| }\Bigg).
\end{equation}

Intuitively, Equation~(\ref{eq:model_equations2}) combines the two constraints 
on the rate at which $i$ upload pieces to $j$. The first term stands for the
maximum instantaneous rate irrespective of capacity limitations. The
second term reflects the fraction of $i$'s uplink capacity that can be   
dedicated to $j$ given that some bandwidth has already been allocated. In this 
case, $c_l - \sum_{k|b_k>b_j}u_{ik}$ is the remaining capacity of $i$ 
and $N-1-|\{k|b_k>b_j,k\neq i\}|$ is the number of peers that will share 
it (including $j$). Note that the equations above relax our initial assumption 
that $b_i,\forall i$ had to be distinct at all times, allowing for leechers to join 
the system simultaneously or more generally, for leechers to have the same number 
of pieces.

In Equations (\ref{eq:bandwidth-requirements2}) and (\ref{eq:model_equations2}),
variables $N$, $b_i$ and $b_j$ change over time, representing arrivals, departures and
the acquiral of new pieces.
Instead of writing those variables as a
function of time, we dropped the $t$ variable for the sake of simplicity.
Therefore, these equations can be computed for each time interval by assigning to
these variables their corresponding values at that time. However, note that
a change in $b_i$ does not necessarily imply in a change in the download rate of 
leechers as what matters is the relationship between $b_i$ and $b_j$, for all $i,j$.
Thus, as the system evolves the variables that govern the equations will change value, 
but not the equations themselves, which can be used to compute the current download 
rate of leechers given the state of the swarm.

%

We will see in Section~\ref{sec:validation} that the 
proposed model given by Equations (\ref{eq:bandwidth-requirements2}) and (\ref{eq:model_equations2}) 
yields accurate results for unpopular swarms, indicating that 
for this scenario it is sufficient to know the number of pieces each peer 
possesses. Nevertheless, we further discuss two useful generalizations to this 
model in Section~\ref{sec:general}.

\subsection{Model Validation}
\label{sec:validation}

Our model gives an approximation to the average download rate experienced by 
a leecher in a unpopular swarm, which depends on the relationship between the number 
of pieces owned by the peers and upload capacities. In this section, we validate the model 
comparing its predictions with simulations results. We will see that even though
the model does not take the TFT and other mechanism into account, its results are 
very similar to those obtained from our simulator, which implements a fully functional 
version of the BT protocol (see simulator description in Section \ref{subsec:strange}).

We consider homogeneous swarms with $c_s = c_l = 64$ kBps, where exactly $N = 5$
arrivals occur.  In addition, all leechers arrive
before the first one completes the download and all the arrivals
occur before any two leechers synchronize their contents. In our simulations, we
say that two leechers $i$ and $j$ are synchronized if they have roughly the same
number of pieces, i.e., $|b_i-b_j| < 3$. We use deterministic arrivals to
reproduce the exact scenarios we intend to compare.
\begin{figure*}[h]
\centerline{\subfloat[Evolution of downloaded
    pieces.]{\includegraphics[width=2.7in]{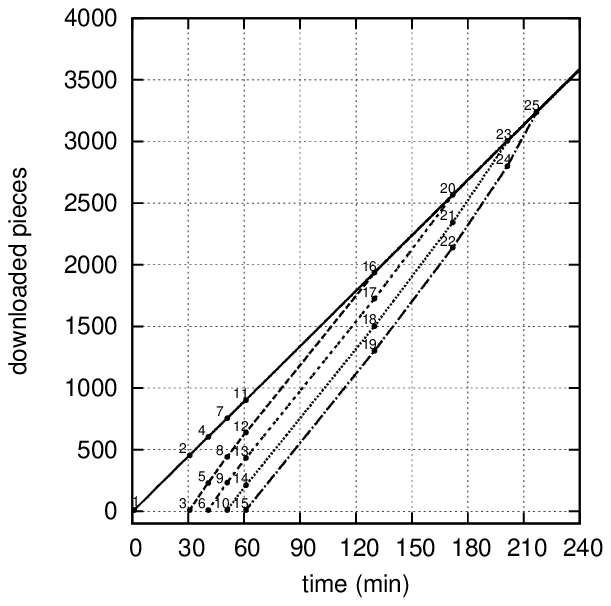}
\label{fig:validation_simul}}
\subfloat[Comparison between simulation and model
    results.]{\includegraphics[width=2.7in]{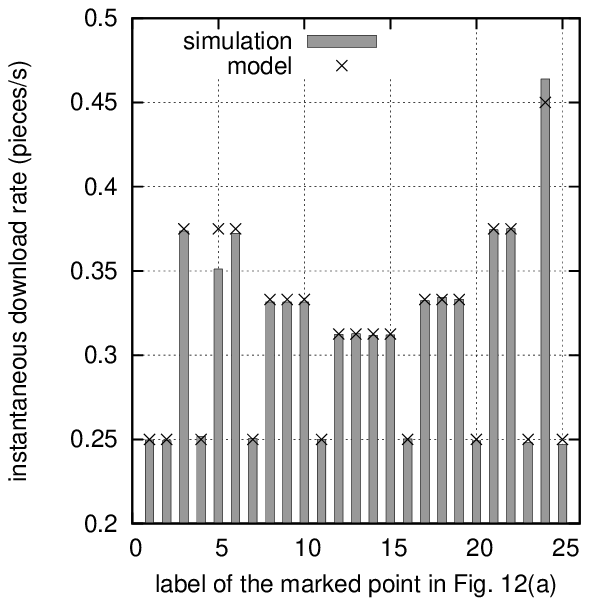}
\label{fig:validation}}}
\caption{Model validation and comparison with simulation.}
\label{fig:comparison}
\end{figure*}

Consider the evolution of number of downloaded pieces in such a swarm
illustrated in Figure~\ref{fig:validation_simul}.  The first leecher arrives at
time $t = 0$ and four other leechers join the swarm at $t = 30,40,50,60$.  After $t =
120$, leechers start to synchronize.  We chose several points from curves in
this figure corresponding to instants of time where an event that can change
peers' download rates occurs.  More precisely, we labeled points in these curves
with numbers when new leechers arrive or when two leechers synchronize.

Figure \ref{fig:validation} shows peers' download rates from simulations and
model for the labeled points indicated in \ref{fig:validation_simul}.  We have simulated
five runs for each scenario including the one depicted in this figure.  The confidence intervals
obtained are relatively small and are omitted.

The simulation results for points 1, 2, 4, 7, 11, 16, 20, 23 and 25 show
approximately the same download rate, what is correctly captured by the model.
The download rates obtained from the model are exactly the same due to fact that
the corresponding peers already have every piece that was previously pushed by
the seed into the swarm. Thus, neighbors of such peer can only upload to it
new pieces they receive directly from the seed, i.e., their upload rate is
constrained by $c_s/N$. Since this constraint is below the capacity that can be
allocated to serve a neighbor when $c_s = c_l$ (which is, at least $c_s/(N-1)$),
every peer in the swarm will upload to one such peer with rate equal
to $c_s/N$. Therefore, the average download rate predicted by the
model for peers in $A$ is $c_s/N + (N-1)c_s/N = c_s = 0.25$. In particular,
the relative error is less than 1.5\% for all these points. The model is quite
accurate even for other values of $N$.

On the other hand, simulation results for the other points exhibit a great
variety of download rates. However, those points which correspond to the same
moment in time display similar download rates (e.g. 8, 9 and 10). We
observe that the download rates decrease with new arrivals. We also note
that as more leechers
become synchronized, non-synchronized leechers achieve higher download rates
(see points 21, 22 and 24). This increase in the download rates occurs
because the greater is the number of synchronized peers, the greater is the
remaining capacity to serve leechers with less pieces. This is due to the fact
that the rate at which synchronized leechers can transmit to each other is very
constrained as we discussed before. The relative error of the model is less than
1\% for all points, but the 5-th (7\%) and the 24-th (3\%).

From these figures we conclude that when $c_s = c_l$, at a given moment in time,
it is possible to partition the set of
leechers in two subsets: leechers with the same number of pieces as the oldest
leecher (subset $A$), and those with less pieces than the oldest one (subset
$B$). When $c_s = c_l$, the model predicts that all leechers in each of
these subsets will have identical download rates. Moreover, a leecher in $B$
will have a higher download rate than one in $A$ and this difference depends on
the set sizes. In particular, larger swarms imply lower values of the minimum
download rate and higher values of the maximum download for
leechers in $B$. This tendency can be observed both in simulation and
model.


Considering all the simulations performed, we conclude
that the model is quite accurate, with differences being unnoticeable in most 
scenarios and less than 10\% in all cases. More importantly, the model captures 
well the trends observed in simulation.

\subsection{Model generalizations}
\label{sec:general}

Some of the assumptions of the model we propose are: (1) unconstrained (or
large) download capacities, and (2) leechers with identical upload
capacities.  We now relax the former assumption by providing an upper bound for
the download rate of a peer. This bound is a function that does not grow fast on
the system parameters.  Clearly, if the download capacities are greater than
this function then all the previously presented results hold.

In what concerns the latter assumption, we indicate how to adapt the model to 
cope with similar (but not identical) upload capacities. Furthermore, we present 
some simulation results that show that the general behavior of the system under 
this scenario is similar to the one presented in Section~\ref{sec:problem}.

\subsubsection{Finite (and small) download capacities}

When the oldest peers are synchronized, they can only send to each other what
they receive directly from the seed (see Equation (\ref{eq:bandwidth-requirement2})). 
This constraint leads to more capacity available to serve those peers that are not 
synchronized. In particular, if there is only one non-synchronized peer, it can benefit from this
idle bandwidth alone and consequently achieve the highest possible download
rate. In what follows we compute an upper bound for this maximum download rate.

Consider an unpopular swarm with $N>1$ peers, such that the $N-1$ oldest peers are
synchronized. From Equation~(\ref{eq:bandwidth-requirements2}) we can compute the maximum instantaneous upload rate of
a synchronized leecher  $i$ to the other peers irrespective of capacity limitations:
\begin{subnumcases}{g_{ij}=}
 \infty,     & if j is not synchronized\\
   \frac{c_s}{N} , & if j is synchronized
  \label{eq:bandwidth-requirement3}
\end{subnumcases}

According to Equation~(\ref{eq:model_equations2}), each leecher $i$ will upload to
each of the other $N-2$ synchronized peers at rate $\min\{\frac{c_l}{N-1}, \frac{c_s}{N}\}$. 
The remaining capacity of $i$ that can be used to serve the younger leecher $N$ is
$c_l - (N-2) \min\{\frac{c_l}{N-1}, \frac{c_s}{N}\}$. Since there are $N-1$ synchronized 
leechers, the capacity that can be used to serve only non-synchronized leecher 
is $(N-1)[c_l - (N-2) \min\{\frac{c_l}{N-1}, \frac{c_s}{N}\}]$. In addition, the 
younger leecher downloads from the seed at rate $\frac{c_s}{N}$. Therefore, the maximum 
download rate is given by
\begin{eqnarray*}
d_{\max} &=& (N-1)c_l - (N-2) \min\{c_l, \frac{c_s(N-1)}{N}\}  + \frac{c_s}{N} \label{eq:equality}
\end{eqnarray*}
Thus, we have:
\begin{subnumcases}{d_{\max}=}
   c_l + \frac{c_s}{N}                     & if $c_l \leq c_s(N-1)/N$ \\
   (c_l - c_s)(N-1) + 2c_s - \frac{c_s}{N} & otherwise
  \label{eq:dmax}
\end{subnumcases}

Note that in both cases the maximum download rate is a value that does not grow fast in any 
of the system parameters. In particular, for small $N$, which is the case of interest, $d_{\max}$ 
has a relative small value with respect to the upload capacities or leechers and seed. Thus, if 
the download capacities of leechers are larger than $d_{\max}$, then results predicted by the model 
are just as good. This condition replaces the requirement of unbounded (or arbitrarily large) 
download capacities assumed earlier in the model. 

To illustrate, consider the example in Section \ref{sec:validation} where $c_s = c_l = 64$ kBps 
and $N=5$. In this case, the highest download rate would be $d_{\max} = 2 \times 64 + 64/5 = 140.8$ kBps. 
Thus, if download capacities of leechers are larger than 140.8Kbps, then the results predicted by the model 
would be unchanged. 

\subsubsection{Similar but not identical upload capacities}

Although we have assumed upload capacities of peers to be identical, this is
certainly not necessary for the piece distribution process in unpopular swarms
to lead to heterogeneous download rates. Note that our modeling framework allows
for peers to have different upload capacities, as $c_l$ could depend on $i$ in
Equation~\ref{eq:model_equations} (equivalently, in Equation~\ref{eq:model_equations2}).
Clearly, this would have an impact on the heterogeneity of the performance and would depend on
the values of $c_{l_i}$ and the order of their arrivals to the swarm.
However, if $c_{l_i}\,\forall i$ are close to one another, for example, chosen uniformly at 
random from a small range, then we expect not to see much differences with respect to 
the constant $c_l$ value. 

\begin{figure}[!t]
\begin{center}
\captionsetup[subfloat]{margin=0.5pt}
\subfloat[$\epsilon=0.25$.]{\label{fig:clvar0.25}
\includegraphics[width=2.8in]{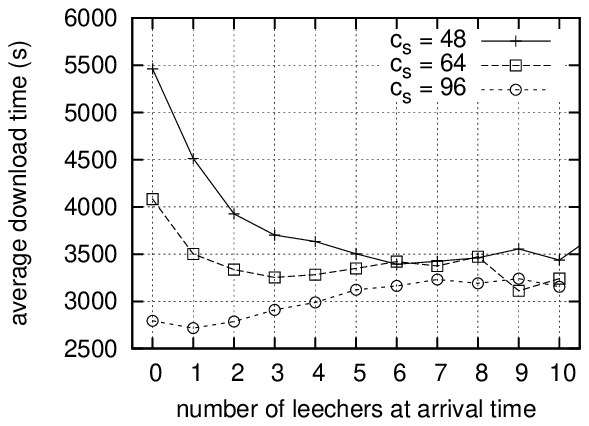}}
\subfloat[$\epsilon=0.50$.]{\label{fig:clvar0.50}
\includegraphics[width=2.8in]{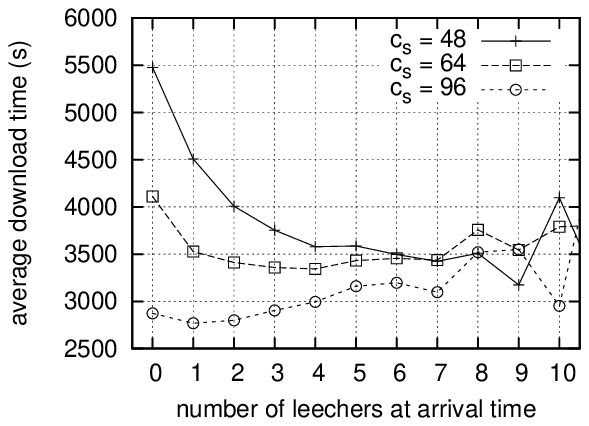}}
\caption{\label{fig:varying-cl} Average download time as a function of arrival
    order in a busy period, when the uplink capacity is $c_{l_i} \sim
        U(c_l(1-\epsilon),c_l(1+\epsilon))$, with $c_l = 64$ kBps}.
\end{center}
\end{figure}

In order to support this last claim, we repeat the simulations described in
Section~\ref{sec:poisson} but allowing the upload capacity of leecher $i$ to be drawn
uniformly at random from the range $[c(1-\epsilon),c(1+\epsilon)]$, where
$c = 64$ kBps and $\epsilon \in \{0.25,0.50\}$.
Figures~\ref{fig:varying-cl}(a-b) show the average download time of peers
binned according to the number of leechers in the swarm at the arrival time for
$\epsilon = 0.25$ and $0.50$, respectively.
We conclude that, when the upload capacities are close to each other,
the system exhibits a very similar behavior to that we observed when the upload
capacities are the same (see Figure~\ref{fig:xx-64_1000_1000_relative-order}).
Not surprisingly, the larger the range of upload capacities, the the greater 
the impact on the results, when compared to a constant upload capacity. 

\section{Predicting bursty departures}
\label{sec:application}

The model presented in Section \ref{sec:model} can be used to estimate the 
number of departures that occur in a burst. In particular, consider the 
arrival of a leecher that initiates a busy period (i.e., the first arrival to 
a swarm with no leechers). In the following, we estimate the average number 
of peers that depart the swarm in a burst together with the leecher that 
initiated the busy period.

In practice, bursty departures do not occur exactly at the same time due to
variations inherent to the network and to the inexistence of mechanisms that
enforce synchronization between peers implemented in the protocol (e.g.: they do
not request pieces at exactly the same time).
Nonetheless, our model does not take
these factors into account and, thus, we focus on leechers that leave the swarm
at exactly the same time as the first leecher.

Let $f$ denote the first leecher of a busy period and assume that the leecher 
arrival follows a Poisson distribution with rate $\lambda$. Also, as assumed by 
the model, a seed is always present and has uplink capacity of $c_s$, while
leechers have identical uplink capacities equal to $c_l$. Finally, 
let $S$ denote the number of pieces of the content.

We know that each leecher downloads pieces from the seed at rate $c_s/N$, where
$N$ is the number of peers in the swarm. These pieces are interesting to all the
other $N-1$ peers and can be sent to them. Thus, if $c_l <
c_s\times\frac{N-1}{N}$, leechers will start to accumulate pieces received from
the seed which cannot be uploaded to the other peers. Therefore, every leecher will own pieces
interesting to all of its neighbors. Consequently, the upload rate between any two
leechers $i$ and $j$ will be equal to $u_{ij} = c_l/(N-1)$, since $g_{ij} =
\infty$ (see Equation~(\ref{eq:model_equations2})). We conclude that when $c_l <
c_s\times\frac{N-1}{N}$, all leechers have the same download rate which prevents
other leechers from departing in a burst with $f$.

Conversely, when $c_l\geq c_s\times\frac{N-1}{N}$, the neighbors of $f$ can
upload
to it the pieces they download from the seed. Since
leecher $f$ downloads from the seed at rate $c_s/N$ and each of its $N-1$
neighbors receives
pieces from the seed and uploads them to $f$ at the same rate, $f$ will download
the content at a constant rate equal to $c_s$,
independently on the number of peers in the swarm. 
Note that $c_s$ is also the upper bound on the average download rate, as the seed 
cannot uploads pieces into the swarm at a faster rate. Hence, leecher $f$ will take 
$T = S/c_s$ seconds to finish the download.

We now show how to calculate the lower and upper bounds to the number of bursty
departures when $c_l \geq c_s\times\frac{N-1}{N}$.
Consider arrivals that occur while peer $f$ is in the swarm. The number of such 
arrivals, say $n=N-1$, is a random variable and follows the Poisson distribution with 
parameters $\lambda$ and $T$. The download rates of these leechers are a function 
of $n$ and also of their instants of arrival. Moreover, as discussed in
Section~\ref{sec:validation}, larger values of $n$ imply a larger spread in the
download rates. To obtain a conservative lower 
and upper bound on these download rates, we will consider a sufficiently large 
value for $n$. In particular, we use the 99-th percentile of $n$, namely $n_{99}$, and 
thus, $P[n \leq n_{99}] \leq 0.99$.

Given that exactly $n_{99}$ leechers will join the swarm before the departure of 
$f$, we can use the model to obtain the minimum and maximum download rates of these 
peers, independently of their inter-arrival timing. Let $d_{\min}$ and $d_{\max}$
be, respectively, the minimum and the maximum download rates obtained from the model
given that the swarm has $n_{99}+1$ leechers.

Consider again the subsets presented in the previous section, namely $A$
(leechers with the same number of pieces than $f$) and $B$ (leechers with less
pieces than $f$). The minimum download rate is obtained by a leecher $m$ in $B$ when
the only leecher in $A$ is $f$. In this case, the download rate $d_{\min}$ is given by
\begin{equation}
d_{\min} = \frac{c_s}{n_{99}+1} + u_{f,m} + \sum_{i \in B} u_{i,m},
\end{equation}
where $\sum_{i \in B} u_{i,m}$ corresponds to the sum of the rates at which $m$
downloads from peers in $B$.

On the other hand, a leecher $m$ in $B$ obtains the maximum download rate
$d_{\max}$ when it is the only peer in $B$, i.e, $|A|=n_{99}$. In this
case, the download rate is given by
\begin{equation}
d_{\max} = \frac{c_s}{n_{99}+1} + \sum_{i \in A} u_{i,m}.
\end{equation}

The minimum and maximum time for 
the leechers to download the content is, respectively, $S/d_{max}$ and $S/d_{min}$.
Therefore, at least all leechers that arrive before $T - S/d_{min}$ will 
leave the swarm together in a burst with $f$. The expected number of peers that
will arrive within this time period, $B_{min}$ is simply given by 
\begin{equation}
B_{min} = \lambda \, \left( T - \frac{S}{d_{min}} \right).
\label{eq:B_min}
\end{equation}

Similarly, at most all leechers that arrive before $T - S / d_{max}$ will 
leave the swarm in a burst with $f$. The expected number of peers that will arrive 
within this time period, $B_{max}$ is simply given by 
\begin{equation}
B_{max} = \lambda \, \left( T-\frac{S}{d_{max}} \right).
\label{eq:B_max}
\end{equation}

Finally, $B_{min}$ and $B_{max}$ provide a lower and upper bound for the average 
number of leechers that will depart the swarm in a burst with $f$. 

\begin{table}[!t]
\centering
\caption{Bounds for the expected number of leechers that depart in a burst with $f$, for $\lambda=1/1000$.}
\label{tab:burst-bounds}
\begin{tabular}{c|c|c|c|c|c}
$c_{s}$ & \multirow{2}{*}{$E[N]$} &  \multirow{2}{*}{$B_{min}$} & \multirow{2}{*}{$B_{max}$}
& \multirow{2}{*}{$\frac{\textrm{$B_{min}$}}{E[N]}$}
& \multirow{2}{*}{$\frac{\textrm{$B_{max}$}}{E[N]}$}\\
(kBps) & & & & &\\
\hline
48 & 5.333 & 1.667 & 4.378 & 0.312 & 0.821\\
64 & 4.000 & 0.400 & 1.895 & 0.100 & 0.474\\
\end{tabular}
\end{table}

Table~\ref{tab:burst-bounds} shows the expected number of arrivals to the swarm 
before $f$ departs, $E[N]$, which is simply $\lambda T$, and both the lower and 
upper bounds $B_{min}$ and $B_{max}$, respectively. The table shows numerical 
results for different $c_s$ values but with $c_l = 64$ kBps and $\lambda = 1/1000$.
The results indicate that average number of peers that depart the swarm in a burst 
with $f$ can be significant: between 31\% and 82\% of all arrivals when the seed 
is slower than the leechers and between 10\% and 47\% when they have the same 
upload capacity. We also observe that these ratios reduce as $c_s$ increases, 
indicating that bursty departures are less likely to occur with faster seeds.
Recall that, as indicated above, there is a minimum value of $c_s$ for which
bursty departures do not occur.


\section{General discussions}
\label{sec:disc}

We now discuss other aspects related to the described phenomenon such as
different arrival processes, what happens if the seed is not available all the
time, what happens when leechers stay as seeds for some
time and the missing piece syndrome.

\subsection{General arrival processes}
It is interesting to consider the occurrence of the observed phenomenon in more 
general scenarios. Although we have shown its prevalence under a crafted peer 
arrival process and under Poisson arrivals, we claim that homogeneous peers 
can have heterogeneous download rates under very general arrival patterns. 
In particular, given any arrival pattern of peers into a swarm, it 
is possible to choose system parameters (i.e., seed upload capacity, leechers upload capacity, 
and file size) such that the effects described in this paper will be very prevalent. Intuitively,
by choosing a fast enough seed, peers will not be able to disseminate old pieces
before new ones are pushed into the swarm, and thus will have 
significantly different number of blocks. In a sense, the 
behavior observed and described in this paper is quite general, although the requirement 
of the swarm being unpopular is important, as we next describe. 

\begin{figure}[!t]
\centering
\captionsetup[subfloat]{margin=0.5pt}
\subfloat[Number of leechers over time]{\label{fig:popular_busyperiod}
\includegraphics[width=2.8in]{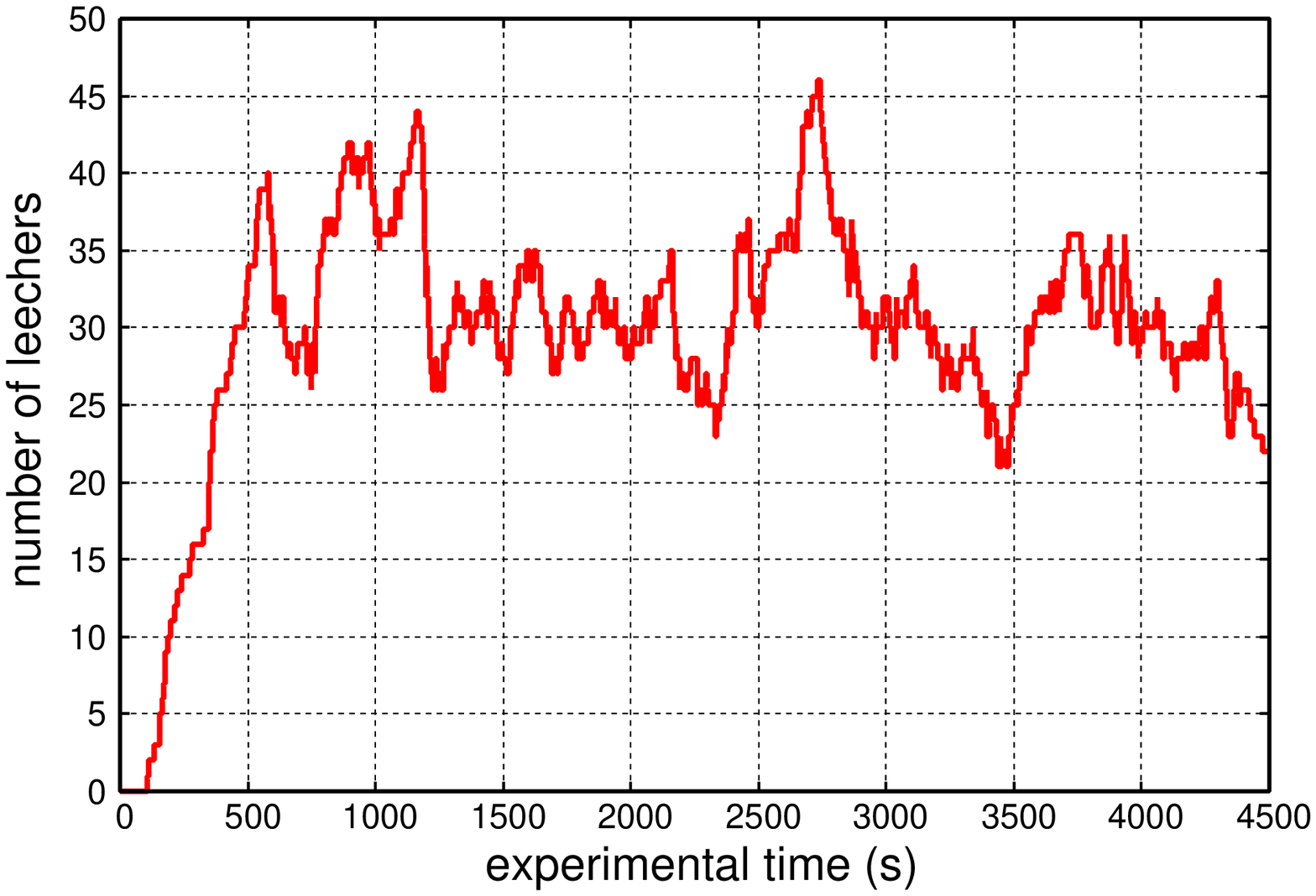}}
\subfloat[Empirical CCDF of the download time.]{\label{fig:popular_CCDF}
\includegraphics[width=2.8in]{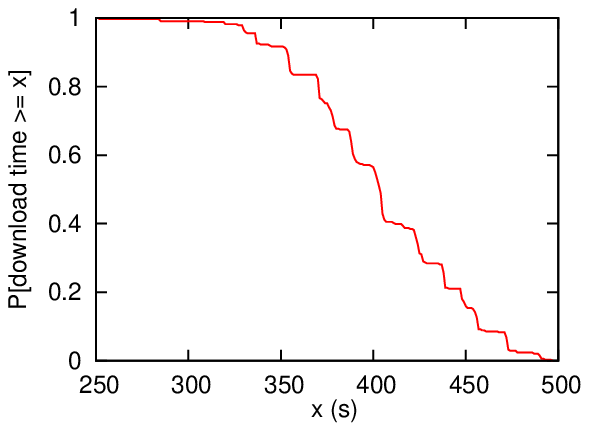}}
\caption{Experimental results under a popular swarm ($\lambda = 1/12$,
$c_s= 50$ kBps, $c_l=50$ kBps)}
\label{fig:popular}
\end{figure}

What happens if we consider very popular swarms, where the peer arrival rate is very large, 
yielding very large swarm sizes? Figure \ref{fig:popular_busyperiod} shows experimental results of the 
dynamics of leecher arrivals and departures for this scenario (Poisson arrivals with 
rate $\lambda = 1/12$, uplink capacities of $c_s = 50$ kBps and $c_l = 50$
kBps) and file size $S=80$ pieces (i.e., 20 MB). The empirical CCDF of the
download time is depicted in Figure~\ref{fig:popular_CCDF}.
Interestingly, we can still observe the consequences of having heterogeneous 
download rates, such as bursty departures, content synchronization 
and high variability of download times (peers that leave in a large burst have different 
download times, as arrival is well-behaved), for example, at times 600 s and 1200 s. In a 
sense, the phenomenon is quite prevalent even during the busy period, but not strong 
enough to end the busy period. The characterization and modeling of the phenomenon in 
this scenario is much more entailed, given the complicated dynamics of piece exchange 
of BT and consequently the interest relationship among peers. We leave the investigation 
of these scenarios (popular swarms) as future work.

\subsection{When the seed is not available all the time}

We have considered so far swarms that have a single seed which is always
connected. However, what happens if the seed alternately joins and leaves the swarm?
Intuitively, leechers start to synchronize their contents right after the seed
leaves because no new pieces are being placed into the swarm. After they become fully
synchronized, they will stall until the seed comes back. Then, since they are
synchronized, they will have relatively low download rates and will leave almost
at the same time. Therefore, the intermitent seed makes the average download rates even 
more heterogeneous.

In order to support this claim, we modify the simulation model such that the state of the 
seed (connected/disconnected) is given by an ON-OFF source. We assume that the time until the seed leaves 
the ON state (leaves the swarm) is exponentially distributed with rate $1/\lambda$ (arbitrarily set to be equal to the 
leecher arrival rate). Furthermore, we choose the rate at which the seed goes from the OFF back to the
ON state (rejoins the swarm) so that the availability of the seed is $0.75$, $0.50$ and $0.25$.

Table~\ref{tab:on-off} summarizes the results for $\lambda=1/1000$ s, $c_s = c_l = 64$
kBps, $S= 1000$ pieces. Each scenario was simulated during $800,000$ s. We
observe that
the mean, variance and maximum download time monotonically increase, what is
expected as there are less resources on average. Interestingly,
the minimum download time becomes smaller. This is due to the fact that a
a new leecher may arrive when some peers are stalling in the absence of the
seed, just before finishing the download.
The new leecher then benefits from the spare bandwidth capacity and
might complete the download right after the seed comes back.
Finally, it is clear the download time (equivalently, download
rate) becomes more heterogeneous.

\begin{table}[ht]
\centering
\begin{tabular}{c|c|c|c|c}
$P(S=\textrm{ON})$ & Mean & Variance & Minimum & Maximum\tabularnewline
\hline
1.00 & $3.360\times10^{3}$ & $9.319\times10^{4}$ & $2.295\times10^{3}$ & $4.247\times10^{3}$\tabularnewline
0.75 & $3.865\times10^{3}$ & $1.003\times10^{6}$ & $1.276\times10^{3}$ & $8.431\times10^{3}$\tabularnewline
0.50 & $5.307\times10^{3}$ & $1.062\times10^{7}$ & $5.640\times10^{3}$ & $2.518\times10^{4}$\tabularnewline
0.25 & $1.045\times10^{4}$ & $7.671\times10^{7}$ & $3.640\times10^{3}$ & $3.321\times10^{4}$\tabularnewline
\end{tabular}
\caption{Statistics of the empirical distribution of the download time, when the
seed leaves and joins the swarm.}
\label{tab:on-off}
\end{table}

Note that, in fact, the proposed analytical model
can cope with this scenario as long as peers do not accumulate pieces
interesting to each other, i.e. $c_l \geq c_s \times \frac{N-1}{N}$ (see
Section~\ref{sec:application}). In particular, if the seed departs, then this
will only affect the upload rate among peers, given by
Equation~(\ref{eq:bandwidth-requirements}) (or
Equation~(\ref{eq:bandwidth-requirements2})).
More precisely, the term corresponding to the download rate from the seed ($c_s/N$) should be
set to zero in the equations that describe time periods where the seed is not
present. Thus, given the state of the swarm with respect to the seed's presence and 
number of leechers, we can still apply our modeling framework and determine 
the download rates of leechers (under the condition above).

\subsection{When leechers stay as seeds for some time}

\begin{figure}[!t]
\begin{center}
\captionsetup[subfloat]{margin=0.1pt}
\subfloat[Bursty departures characterization.]{\label{fig:bursty_departures-seeding}
\includegraphics[width=2.2in]{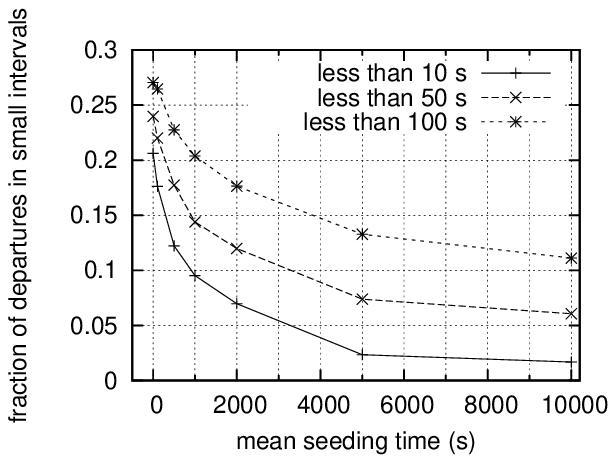}}
\subfloat[Empirical CCDF of the download time.]{\label{fig:download_times-seeding}
\includegraphics[width=2.2in]{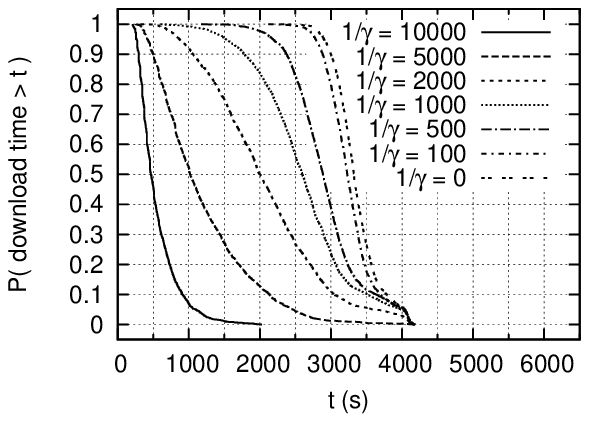}}
\subfloat[Impact of arrival order.]{\label{fig:relative_order-seeding}
\includegraphics[width=2.2in]{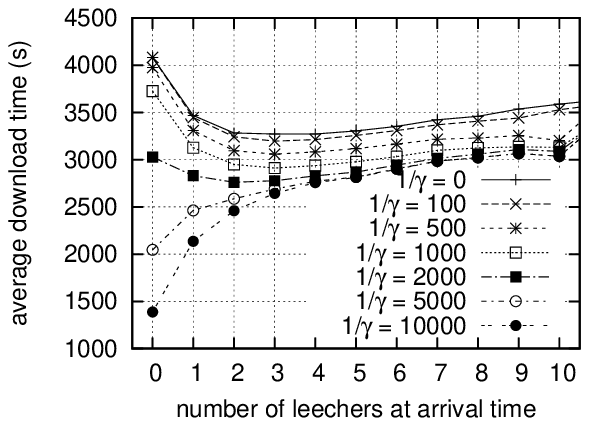}}
\caption{\label{fig:seeding} System performance when leechers stay seeding.}
\end{center}
\end{figure}

Another aspect that can be taken into account is that leechers may stay as seeds
for a period of time after they finish the download and before leaving the
swarm. Intuitively, since the capacity available to disseminate the file increases as
leechers stay as seeds, peers concurrently downloading a file tend to receive
pieces at similar download rates, possibly reducing the consequences of
heterogeneous download rates.
We performed
simulations for scenarios where $\lambda = 1/1000$ s, $c_s = c_l = 64$ kBps, $S
= 1000$ pieces and the time which leechers stay seeding is deterministic and
equal to $1/\gamma$. Each scenario was simulated 10 times during 400,000~s,
but the first 100,000~s were discarded to avoid transient effects.
Figures~\ref{fig:seeding}(a)-(c) depict the results for many values
of $1/\gamma$.

As indicated in Figure~\ref{fig:bursty_departures-seeding}, bursty departures
are less likely to occur when leechers stay in the swarm after downloading the
entire content. However, for small values of $1/\gamma$ (with respect to
$1/\lambda$), the difference is barely noticeable and the departure
process is still very bursty.

Figure~\ref{fig:download_times-seeding} shows the
CCDF of leechers' download times for different values of $1/\gamma$ while
Table~\ref{tab:seeding} contains statistics of these distributions, namely the
sample mean and variance, minimum and maximum values.  Intuitively, leechers
that find two or more seeds at arrival time have significant better performance
and hence the minimum download time decreases as the seeding time increase. On
the other hand, the maximum download time is the approximately same for the majority of the
curves. This is because there is a non-zero probability that a leecher downloads
the content entirely from a single seed. However, this probability becomes
smaller with the seeding time. Initially the variance increases with $1/\gamma$, but
when leechers stay as seeds for a long period of time,
it is unlikely that a leecher will have a download time much larger than the
average and thus, the sample variance diminishes (see $1/\gamma = $ 5,000 and
$1/\gamma = $ 10,000 in Table \ref{tab:seeding}).

Finally,
Figure~\ref{fig:relative_order-seeding} shows that while early arrivals are
detrimented for small values of $1/\gamma$, they are benefited when
$1/\gamma$ is high. The presence of multiple seeds has the same effect as a
single seed with higher capacity.  This can be observed by comparing
the curves for $1/\gamma$ equal to 5,000 and 10,000 and the
curve corresponding to $c_s = 96$ kBps in
Figure~\ref{fig:xx-64_1000_1000_relative-order}.

\begin{table}[!t]
\centering
\caption{Statistics of the empirical distribution of the download time, when
    leechers stay as seeds for some time.}
\label{tab:seeding}
\begin{tabular}{c|c|c|c|c}
$1/\gamma$ & Mean  & Variance & Minimum & Maximum\\
\hline
0 &     $3.360 \times 10^3$  & $9.319 \times 10^4$ & $2.295 \times 10^3$ & $4.247 \times 10^3$\\
100 &   $3.262 \times 10^3$  & $1.093 \times 10^5$ & $2.120 \times 10^3$ & $4.181 \times 10^3$\\
500 &   $2.926 \times 10^3$  & $2.446 \times 10^5$ & $1.174 \times 10^3$ & $4.183 \times 10^3$\\
1000 &  $2.612 \times 10^3$  & $4.394 \times 10^5$ & $6.028 \times 10^2$ & $4.181 \times 10^3$\\
2000 &  $2.067 \times 10^3$  & $6.278 \times 10^5$ & $3.197 \times 10^2$ & $4.170 \times 10^3$\\
5000 &  $1.190 \times 10^3$  & $4.306 \times 10^5$ & $2.600 \times 10^2$ & $4.209 \times 10^3$\\
10000 & $5.508 \times 10^2$  & $7.614 \times 10^4$ & $2.100 \times 10^2$ & $2.028 \times 10^3$\\
\end{tabular}
\end{table}

\subsection{Missing piece syndrome}

Last, we now comment on the relationship of our findings and the phenomenon
known as {\it missing piece syndrome}. This phenomenon states that in swarms
where the arrival rate is large enough, the 
system can become unstable (i.e., number of leechers grows unboundedly) if the upload 
capacity of the seed is not large enough \cite{mathieu2,hajek_zhu_2010,hajek2}.
The key aspect of this syndrome is content 
synchronization, where a large fraction of peers have all but one and the same 
piece. This situation is particularly bad to the performance of the swarm, as the 
departure rate of the swarm will be equal to the seed upload capacity 
(assuming peers depart as soon as they acquire the last block). Our work has 
shown that peers can synchronize their content in such a way that several
identical pieces are missing which eventually leads to the missing piece
syndrome. In 
some sense, this generalizes the syndrome to a {\it piece synchronization 
syndrome}, which is inherent to BT dynamics due to the heterogeneous download 
rates as discussed in this work. Once peers have synchronized their content, they can only acquire new pieces 
from the seed, at the upload capacity of the seed. In this scenario, 
the {\it missing piece syndrome} is bound to occur.

%
%
%
%
%

\section{Related prior works}
\label{sec:related}


Modeling P2P file sharing systems and in particular BT has been an active area 
of research in the past few years, driven mainly by the high complexity, robustness 
and user-level performance of such systems. One of the first BT models to 
predict the download times of peers was presented in \cite{qiu_srikant_2004}. This 
simple fluid model based on differential equations assumes homogeneous peer 
population (with respect to download and upload capacities) and Poisson arrivals, 
but yields analytical steady state solution for performance metrics. Several subsequent BT models have been 
proposed in the literature to capture various system characteristics, among them 
heterogeneous peer population (with respect to upload and download capacities) 
\cite{piccolo_neglia_2004,liao_papadopoulos_psounis_2007,meulpolder_2008,chow_golubchik_misra_2009}.
However, to the best of our knowledge, all models predict that identical peers 
(with respect to their upload capacities) simultaneously downloading a file will
have similar performance (with respect 
to download rates), contrary to the findings in this paper. Moreover, BT models 
generally assume either a rather large peer arrival rate (e.g., Poisson) or a large 
flash crowd (all peers join the swarm at the same time). This is somewhat surprising, 
given that most real BT swarms are rather small in size and quite unpopular
\cite{guo}. Finally, one perverse effect of this lack of popularity, known as content
unavailability, is shown to be a severe problem found in most of BT swarms~\cite{conext2009}.

Another interesting aspect of BT has been the discovery and characterization of 
some non-trivial phenomena induced by its complex dynamics. For example, peers in BT 
swarm tend to form clusters based on their upload link capacities, exhibiting a strong 
homophily effect. In particular, peers with identical upload capacities tend to exchange 
relatively more data between them
\cite{legout_liogkas_kohler_zhang_2007,bharambe_herley_padmanabhan_2006,performance2010}. 
Yet another peculiar behavior is the fact that arriving leechers can continue to 
download the entire content despite the presence of any seed in the swarm, a property 
known as self-sustainability \cite{menasche_et_al_2010}.
More recently, the {\it missing piece syndrome} has been 
characterized mathematically \cite{hajek_zhu_2010,hajek2}. In
\cite{stability} is presented an evaluation for the impact of different peer
selection strategies on the stability of the system. Such strategies may
reduce the effect of content synchronization among peers.
However, to the best 
of our knowledge, we are not aware of any prior work that has alluded the phenomenon 
we describe in this paper, namely, that homogeneous peers can have heterogeneous 
download rates.

\section{Conclusion}
\label{sec:conc}

This paper identifies, characterizes and models an interesting phenomenon in 
BT: homogeneous peers (with respect to their upload capacity) experience 
heterogeneous download rates. The behavior is pronounced in 
unpopular swarms (few leechers) and has important consequences that directly 
impact peer and system performance. The mathematical model proposed captures 
well these heterogeneous download rates of peers and 
provides fundamental insights into the root cause of the phenomenon. Namely, 
the allocation of system capacity (aggregate uplink capacity of all peers) among 
leechers depends on the piece interest relationship among peers, which for 
unpopular swarms is directly related to arrival order of peers and can be significantly 
different among them.

\section*{Acknowledgment}
This research was supported in part by grants from CAPES, CNPq
FAPERJ (Brazil) and NSF under the grant CNS-1065133.

\bibliographystyle{elsarticle-num}
\bibliography{SyncBT}

\end{document}